\documentclass[letterpaper]{article}

\usepackage[T1]{fontenc}

\usepackage{geometry}
\usepackage{setspace}
\usepackage{algorithm2e}
\usepackage[
  backend=biber,
  style=chem-acs,
  articletitle=true,
  chaptertitle=true,
]{biblatex}
\usepackage{graphicx}
\usepackage{float}
\newfloat{scheme}{htbp}{los}
\floatname{scheme}{Scheme}
\floatname{chart}{Chart}
\newfloat{graph}{htbp}{loh}

\usepackage{chemformula} 
\usepackage[version = 4]{mhchem} 

\usepackage{lineno}
\usepackage{setspace}

\usepackage{authblk}
\author{Soniya Kumawat}
\author{Sayeedul Islam Sheikh}
\author{Satyabrata Patro}
\author{Vishal Singh}
\author{Anurag Tripathi*}
\affil{Department of Chemical Engineering, Indian Institute of Technology Kanpur, 208016, India}

\title{Intercoupling of Segregation and Rheology in Spatially Developing Granular Chute Flows}
\date{*Email: anuragt@iitk.ac.in}

\begin{document}

\maketitle

\begin{abstract}
We investigate the flow and segregation of binary granular mixtures with density differences down a long chute using a continuum framework that couples a particle-force-based segregation model with an inertial-number-based \textcolor{black}{local} rheology. The steady-state momentum and convection-diffusion-segregation equations are solved simultaneously, explicitly accounting for the two-way coupling between segregation and flow. Predicted concentration and velocity fields at different streamwise locations are validated against \textcolor{black}{representative} DEM simulations. The validated model is then used to examine the influence of density ratio, mixture composition, and chute inclination on segregation over a wide range of conditions. The chute length required to achieve fully developed segregation is quantified and compared with the development length for monodisperse granular flow. At low density ratios and/or \textcolor{black}{low} inclinations, segregation develops over much longer distances than the velocity field. In contrast, at higher inclinations and larger density contrasts, the two length scales become comparable, demonstrating that neglecting flow development can significantly underestimate segregation evolution.

\end{abstract}

\section*{Keywords}
Granular materials, Segregation, Continuum model, Discrete Element Method, Chute flow



\section{Introduction}
Granular mixtures rarely remain homogeneous during flow and handling, instead they exhibit segregation where particles separate according to differences in their size, density, and surface properties. A large body of literature addresses granular segregation, and different flow configurations \textcolor{black}{have} been employed to investigate the segregation behaviour of granular mixtures including vibrated bed \parencite{knight1993vibration,rosato2002perspective,menbari2020studying,qiao2021study,dai2024investigation,yang2015confinement,cano2017segregation,mahdi2020effects}, hopper flow \parencite{samadani1999size,ketterhagen2007granular,ketterhagen2009predicting,garcia2016segregation,marucci2018sifting,xiao2019continuum,zhang2020segregation,deng2020modeling}, chute flow~\parencite{dolgunin1998development,savage1988particle,vallance2000particle,kessler2009simulation,hajra2012granular,bhattacharya2014chute}, heap flow \parencite{williams1968mixing,baxter1998stratification,fan2014modelling,schlick2015granular,xiao2016modelling,deng2019modeling,jones2018asymmetric,Duan2021,duan2022designing}, rotating tumbler \parencite{savage1988particle,hill1997axial,zik1994rotational,khakhar2003segregation,santos2016segregation,liao2019effect,chung2021experimental,liao2025experimental,windows2016understanding,yang2017segregation,yari2020size,yang2021continuum}, and plane or annular shear flow \parencite{scott1975interparticle,hill2008isolating,golick2009mixing,may2010shear,TIRAPELLE20211annular,fan2010shear,fan2011theory,Fan2015,fry2018effect,fry2019,cui2022particle,guillard2016scaling,trewhela2021experimental,duan2024general,trewhela2024segregation}. Among the various flow geometries used to study granular segregation, chute flows are mainly employed because of their simple setup and their relevance in both industrial and natural flow situations. Experimental studies~\parencite{dolgunin1995segregation,dolgunin1998development,savage1988particle,vallance2000particle,kessler2009simulation,hajra2012granular,bhattacharya2014chute} on chute flows have demonstrated that differences in particle size or density can cause pronounced segregation. The observations indicate that in dense sheared flows, gravity driven percolation is the dominant segregation mechanism, and shear-induced effects become more significant only when gravity driven segregation is weak or absent.

A large body of literature assumed or prescribed the velocity field to study the evolution of segregation.  \textcite{dolgunin1998development} investigated the segregation of binary mixtures of different sizes and density using experiments and the continuum approach utilizing a known velocity profile.
\textcite{Wiederseiner2011} conducted experiments for binary size mixtures flowing down a long chute and predicted the concentration profile by solving the convection-diffusion equation by incorporating the segregation model proposed by \textcite{grayandchugunov2006particle}. The authors found that the predictions match well in the steady, fully developed region but failed to predict the segregation behaviour in the developing zone. \textcite{zhou2016effect} performed DEM simulations in a long chute to investigate the influence of base roughness on size segregation and flow behaviour.
The results revealed that increasing base friction enhances the rate of size segregation. However, when the base friction coefficient exceeds the particle-particle friction coefficient, the base friction no longer affects the segregation dynamics. \textcite{deng2018continuum} investigated ternary size segregation in long chute flows using both DEM simulations and a continuum model based on an empirical segregation model of \textcite{fan2014modelling} along with a known velocity profile. They showed that the steady-state predictions of the continuum model match well with the DEM data. Their study highlighted the interplay among advection, diffusion, and segregation in governing the segregation dynamics. Furthermore, the continuum model successfully captured the influence of the velocity profile and basal slip conditions on the segregation behaviour. 

A parallel line of work has focused specifically on the streamwise evolution of velocity and depth profiles on which segregation depends. \textcite{ancey2001dry} studied flow behaviour of dry granular materials over rough chute using experiments and theory and \textcolor{black}{observed different types of instabilities and transitions along with flow patterns} that primarily depend on the inclination angle. In particular, for inclination angles below the material's internal friction angle, the flow becomes unstable, resulting in the formation of a stationary wedge along the chute bed. \textcolor{black}{Later work by \textcite{taberlet2003superstable,taberlet2004growth} showed that sidewalls-stabilized-heap can maintain a constant surface inclination while supporting a flowing layer of nearly uniform thickness along its free surface.} \textcite{kumaran2014dense} developed a shallow-layer continuum model for dense granular flows based on asymptotic analysis and dense granular rheology. The model successfully predicted the evolution of flow depth and velocity in long inclined flows. Furthermore, the study investigated the development length and its dependence on the slip parameter, highlighting the role of basal slip in determining the attainment of a fully developed flow. We revisit this dependence for the coupled case of flow and segregation development in section~\ref{subsec:effect_density_comp_theta}. \textcolor{black}{Recently,} \textcite{wu2025unified} performed discrete element method (DEM) simulations to investigate the evolution of velocity profile from the inlet region to the fully developed state. \textcolor{black}{Using different data sets available in literature as well as their own experiments and simulations,} the authors proposed a unified flow rule for dense granular flows down rough inclines to describe both undeveloped and fully developed regimes.

Although extensive experimental studies have investigated segregation and flow behaviour in chute flows, comparatively few studies have employed DEM simulations because of the large system sizes required for chute geometries. To mitigate the high computational cost, most researchers~\parencite{tripathi2013density,tripathi2021size,tirapelle2022cfd,jiang2023rheology,liuandhennan2023coupled,sahu_kumawat_agrawal_tripathi_2023,singhandhennan2024continuum} utilized periodic boundary conditions in their DEM simulations, which allow the study of bulk flow and size-driven or density-driven segregation dynamics without simulating the entire chute length. Continuum models based on empirical segregation models~\parencite{fan2014modelling,xiao2016modelling,jones2018asymmetric,Duan2021} have been primarily validated against bounded heap and rotating cylinder geometries. However, their applicability in chute flows has been questioned due to their failure to predict the effect of inclination angle on segregation~\cite{tripathi2017size}. In parallel, the particle-force-based continuum models have been shown to predict the steady as well as time dependent segregation very well for both size difference~\parencite{tripathi2021size,kumawat2025sizetransient} and density difference~\parencite{tripathi2013density,sahu_kumawat_agrawal_tripathi_2023,kumawat2025transient} but only within periodic chute flows. Imposing periodic boundary conditions in the streamwise direction represents an approximation, since it suppresses the explicit spatial evolution of the flow from inlet to outlet. These studies have further shown that rheology and segregation are intercoupled with each other during transient evolution of flow of granular mixtures in periodic chutes. Whether an analogous coupling governs the spatial development of segregation and flow along a non-periodic chute, however, remains unexplored. In such long chutes, gradients in streamwise direction may also be present. \textcolor{black}{The presence of sidewalls may form the sidewall-stabilized heap and different rheoloy.}

Therefore, in this work, we explore the applicability of particle-force-based segregation model within the convection-diffusion-segregation \textcolor{black}{framework} to predict the evolution of species concentration along the chute length. The present study focuses specifically on density segregation of equal-sized particles. The $\mu(I)$ rheological framework~\cite{tripathi2013density,jop2006constitutivenature}, relating the local stress to a dimensionless inertial number is utilized in this study to predict the streamwise development of velocity and flow depth. \textcolor{black}{Since the presence of frictional side walls lead to more complex flow kinematics, including regions of very low inertial numbers, the applicability of local $\mu-I$ rheology becomes questionable for such cases. Hence, we restrict this study to long chutes without any side walls.} In order to check the validity of particle-force-based model to predict segregation in long chutes, we perform the DEM simulation over a long chute without side walls and compare the spatial evolution at steady-state with the DEM data for a few cases. We then explore the effect of mixture composition, density ratio, inclination angle and basal slip on the extent of segregation using our continuum model. We also investigate the the minimum chute length required to achieve a fully developed flow and segregation and explore the importance of intercoupling of rheology and segregation to accurately predict segregation evolution over a long chute.

The paper is organised as follows.  \S\ref{sec:DEM_simulation} describes the discrete element method (DEM) simulation methodology employed in this work.
The governing equations incorporating the particle-force-based segregation model and mixture rheological model along with the numerical method are described in \S\ref{sec:continuum_model}. The results for \textcolor{black}{developing flows of} monodisperse \textcolor{black}{grains} and binary mixtures comprising particles of different densities are presented and discussed in \S\ref{sec:results}. Finally, the key findings and conclusions of the study are summarized in \S\ref{sec:conlusions}.

\section{Simulation Methodology}
\label{sec:DEM_simulation}
\begin{figure}[ht]
    \centering
\includegraphics[scale=0.5]{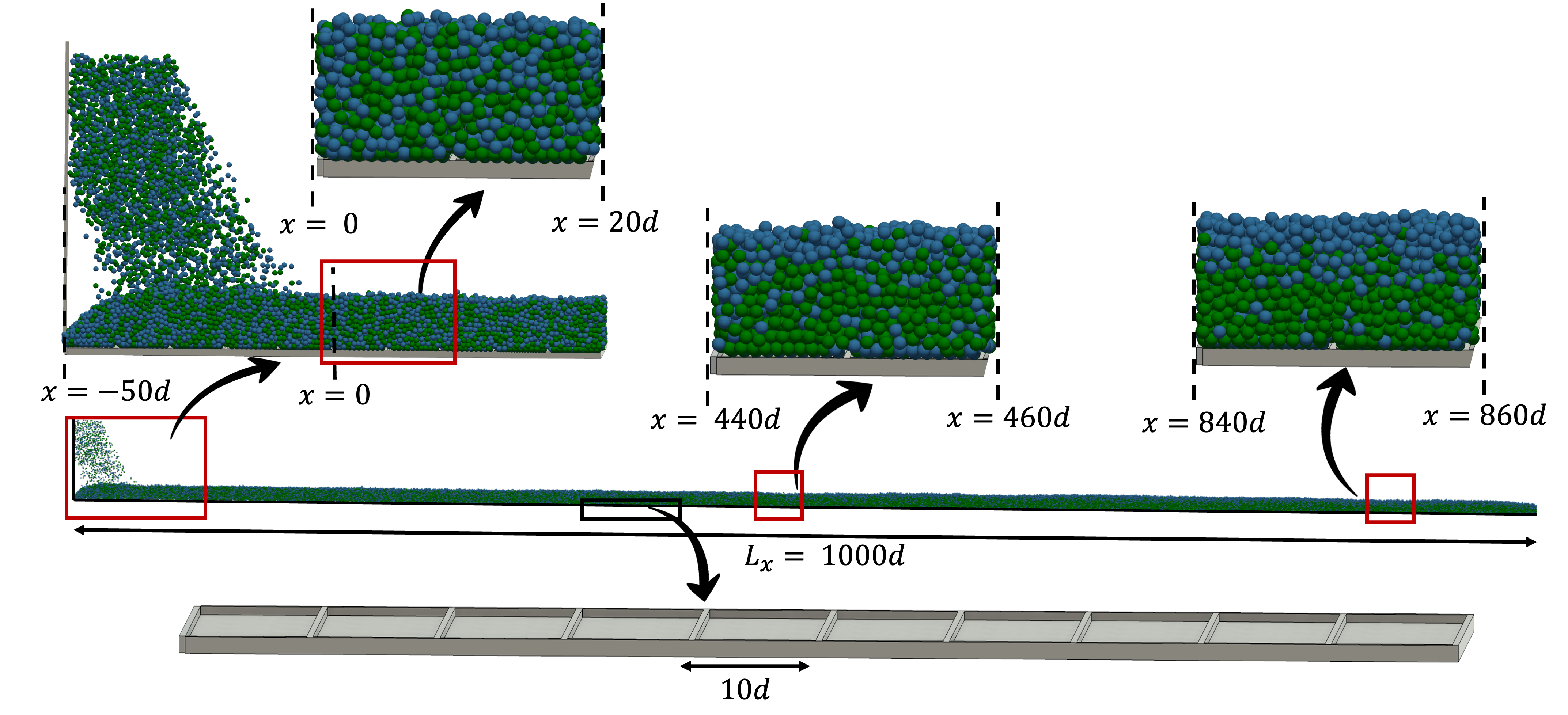}
\caption{Steady-state DEM snapshot of binary mixture of density ratio $\rho = 2.0$ flowing over inclined plane at inclination angle $\theta = 25^\circ$ at different positions. Green spheres represent high density particles while blue spheres correspond to low density particles. \textcolor{black}{Chute base \textcolor{black}{consists of} baffles at \textcolor{black}{regular} intervals of $\Delta x = 10d$ to avoid slip at the base. Segregation evolution along the flow can be observed from the zoomed in view of grains at different $x$ locations.}}
\label{fig:DEM_snaps_theta_29_rho_2.0}
\end{figure}

We performed three dimensional discrete element method (DEM) simulations using the open-source DEM package~\textsc{MUSEN}~\cite{skorych2022parallel} to study the gravity-driven flow of slightly inelastic, frictional particles of identical size. We use same density particles to investigate the \textcolor{black}{developing} flow \textcolor{black}{of grains} and use two different density particles to study the segregation phenomena. In order to study the spatial evolution of \textcolor{black}{properties} along the flow direction, we simulate a \textcolor{black}{thousand-particle-diameter-long chute. As} shown in figure~\ref{fig:DEM_snaps_theta_29_rho_2.0}, the simulation domain is significantly 
elongated in the $x$-direction, with a total length $L_x = 1000d$, and the spanwise width of the domain is $L_z = 20d$. Periodic boundary condition is imposed along the $z$-direction to eliminate the side wall effects. The base of the simulation domain \textcolor{black}{located at $y = 0$} is constructed using baffles of height $0.5d$ at regular intervals of $\Delta x = 10d$ to minimize slip at the base. \textcolor{black}{A fixed wall is imposed at the upstream end (shown as a vertical line at $x = -50d$) in the streamwise $(x)$ direction to prevent particle loss from the left end of the chute. A free outlet condition at the downstream end ($x = 950d$) is implicitly applied \textcolor{black}{in DEM since} the particles leaving the simulation domain are no longer tracked in the simulation.} To maintain a continuous supply of grains, particles are dynamically generated at the upstream region of the incline plane in a \textcolor{black}{particle} generation volume. This \textcolor{black}{cuboidal volume} spans $0 \leq x \leq 20d$ in flow ($x$) direction, $30d \leq y \leq 55d$ in vertical ($y$) direction and full width of the domain in the spanwise ($z$) direction. The box is bounded by \textcolor{black}{flat} vertical walls at $x = 0$ and $x = 20d$. The inclination of the chute is implemented by tilting the gravity vector \textcolor{black}{by the desired angle to get the gravitational acceleration $\vec{g} = g \sin\theta \hat{i} - g \cos\theta \hat{j}$}.

We model the grains as soft deformable spheres of diameter $d = 1\,\mathrm{mm}$ with Young’s modulus $Y = 2.5\times 10^{7}\,\mathrm{Pa}$ and Poisson’s ratio $\nu = 0.25$. The interparticle friction coefficient $\mu = 0.5$ and the coefficient of normal restitution $e = 0.5$ are used to mimic the inelastic and frictional nature of grains. We use Hertz-Mindlin contact force model to compute particle-particle and particle-wall interaction forces in both normal and tangential directions. The expression of contact force for the Hertz-Mindlin model is given as
\begin{equation}
\begin{aligned}
\mathbf{F}_{ij} &= 
\left( 
\frac{4}{3} Y_{\mathrm{eff}} r_{\mathrm{eff}}^{1/2}\,\delta_n^{3/2}
+ 2 \sqrt{\frac{5}{3}}\, \alpha\, (m_{\mathrm{eff}} 
Y_{\mathrm{eff}})^{1/2} (r_{\mathrm{eff}} \delta_n)^{1/4} v_{n,ij}
\right) \, \hat{n}_{ij} \\
&\quad +
\left( 
8 G_{\mathrm{eff}} r_{\mathrm{eff}}^{1/2} \, \delta_n^{1/2} \, \delta_t
+ 4 \sqrt{\frac{5}{3}}\, \alpha\, (m_{\mathrm{eff}} 
G_{\mathrm{eff}})^{1/2} (r_{\mathrm{eff}} \, \delta_n)^{1/4} v_{t,ij}
\right) \, \hat{t}_{ij}
\end{aligned}
\end{equation}
The first two terms contribute to the normal force $\mathbf{F}_{n,ij}$, and the last two terms contribute to the tangential force $\mathbf{F}_{t,ij}$. The quantity $\delta_n$ is the normal overlap and $\hat{n}_{ij}$ is the unit vector in the normal direction. The effective Young’s modulus $Y_{\mathrm{eff}}$ is computed as $\frac{1}{Y_{\mathrm{eff}}} = \frac{1-\nu_i^2}{Y_i} + \frac{1-\nu_j^2}{Y_j}$, where $\nu_{i(j)}$ is the Poisson’s ratio of particle $i(j)$. The effective radius $r_{\mathrm{eff}}$ of the contacting pair is given by $\frac{1}{r_{\mathrm{eff}}} = \frac{1}{r_i} + \frac{1}{r_j}$. The damping parameter $\alpha$ used in the contact force model is related to the coefficient of restitution $e_n$ as $\alpha= \dfrac{ln {(e_n)}}{\sqrt{\pi^2 + (ln(e_n))^2}}$. The term $v_{n,ij}$ denotes the magnitude of the relative normal impact velocity, and $m_{\mathrm{eff}}=\frac{m_i m_j}{m_i + m_j}$ is the effective mass of the contacting pair. Similarly, $\delta_t$ is the tangential overlap and $\hat{t}_{ij}$ is the unit vector in the tangential direction. The shear modulus $G_{eff}$ and Young’s modulus $Y_{eff}$ are related through $G_{eff} = \frac{Y_{eff}}{2(1+\nu)}$ and \textcolor{black}{the effective shear modulus $G_{\mathrm{eff}}$ is defined as $\frac{1}{G_{\mathrm{eff}}}
= \frac{(2-\nu_i)}{G_i}
+ \frac{(2-\nu_j)}{G_j}$}. The quantity $v_{t,ij}$ represents the tangential component of the relative velocity at the contact point. The tangential force is limited by Coulomb’s friction law
$| \mathbf{F}_{t,ij} | \leq \mu \, | \mathbf{F}_{n,ij} |$. The simulation time step $\Delta t$ is chosen to satisfy $\Delta t \leq 0.15\, t_{\mathrm{Rayleigh}}$, where $t_{\mathrm{Rayleigh}} = \left( \pi r_{eff} \sqrt{\frac{2\rho(1+\nu)} {Y_{eff}\left(0.1631\nu+0.8766\right)}} \right)$.

The flow properties such as solids fraction, velocity, and species concentrations are calculated by dividing the simulation domain into a number of spatially uniform bins
\textcolor{black}{along the $x$, $y$ and $z$ directions, with each bin having dimensions $\Delta x$, $\Delta y$ and $\Delta z$.} Solids fraction of $j^{th}$ species in the mixture is calculated as $\phi_j = (1/V) \displaystyle \sum_i w_{j,i} v_{p,i}$, where $w_{j,i}$ is the volume fraction of $i^{th}$ particle of the $j^{th}$ species having volume $v_{p,i}$ in the sampling volume $V$. The total solids fraction of the mixture is obtained as $\phi =\displaystyle \sum_j \phi_j$. The local concentration of $j^{th}$ species at any location is obtained as $f_j =  \frac{\phi_j}{\phi}$. The local mixture velocity of all the $N$ particles, partially or completely situated in the sampling volume $V$, is calculated as $\vec{v}=\displaystyle \sum_{k=1}^{N} w_k \vec {c}_{k}$ where $w_k$ is the volume fraction of particle $k$ in the sampling volume and $\vec {c}_{k}$ is the instantaneous velocity of particle. In order to calculate the partial volume of any particle in a bin, we utilize the coarse-graining strategy proposed by \textcite{weinhart2016influence} \textcolor{black}{and use gaussian distribution function} with coarse-graining width of half particle diameter and cut off distance of three particle diameters. \textcolor{black}{Details of coarse-graining for obtaining continuum properties from particle level properties are listed in supplementary material~\ref{sec:coarse_graining}.} \textcolor{black}{In order to explore the segregation evolution with time, we calculate the vertical center of mass position of the species of interest at any given location $x$ as $y_{j,com} = \frac{\displaystyle \sum_i^{N_{j}} y_{j,i} m_{j,i}}{\displaystyle \sum_i^{N_{j}} m_{j,i}}$, where \textcolor{black}{$m_{j,i}$ and $y_{j,i}$ are the mass and vertical position of \textcolor{black}{center of} $i^{th}$ particle of $j^{th}$ species, respectively \textcolor{black}{and $N_j$ is the total number of particles of $j^{th}$ species with in the region $x$ \& $x + \Delta x$ across the entire layer height and width}.}} 

\section{Continuum model}
\label{sec:continuum_model}
\subsection{Rheology and Momentum balance equations}
\begin{figure}[H]
    \centering
     \includegraphics[scale=0.6, trim=0 0 0 0, clip]{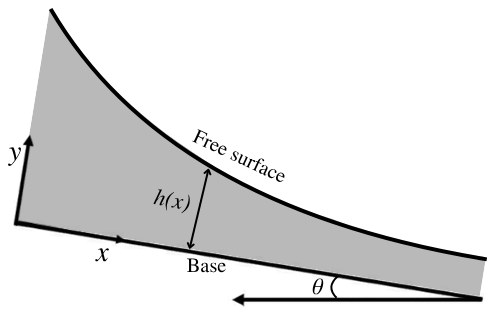}
    \caption{Schematic representation of the spatial flow evolution of a granular flow over a long inclined chute. The shaded region represents the flowing granular layer, and $h(x)$ denotes the local flow depth, which varies with the streamwise coordinate $x$.}
    \label{fig:continuum_model_sketch}
\end{figure}
We consider a thin layer of granular material flowing down an inclined plane at an inclination angle $\theta$. \textcolor{black}{As the flow evolves, the average velocity increases while the layer height decreases along the flow direction. A schematic of the flow is shown in figure~\ref{fig:continuum_model_sketch}.} Due to the periodic boundary conditions applied in the spanwise ($z-$) direction in the DEM simulations, all spatial derivatives with respect to $z$ are taken to be zero in the continuum model. 
Assuming a steady-state flow, the mass balance for this case is written as 
\begin{equation}
    \frac{\partial (\rho_b v_x)}{\partial x} +  \frac{\partial (\rho_b v_y)}{\partial y} = 0.
\end{equation}
The solids fraction of the granular material changes by a small amount \textcolor{black}{along the length of chute for} the range of the inclination angles considered in this study. \textcolor{black}{Hence, we assume \textcolor{black}{the local solids fraction $\phi$} to be constant in our continuum model, \textcolor{black}{giving a constant} bulk density $\rho_b = \phi \rho_p$, $\rho_{p}$ being the particle density. Using this incompressible flow assumption,} the continuity equation reduces to $\frac{\partial v_x}{\partial x} +  \frac{\partial  v_y}{\partial y} = 0$. \textcolor{black}{A simple order of magnitude estimates reveal that} $\frac{\partial v_x}{\partial x}  \sim \frac{v_{x0}}{L}$ and $\frac{\partial v_y}{\partial y}  \sim \frac{v_{y0}}{h}$. \textcolor{black}{Using continuity equation,} we get $v_{y0} \sim v_{x0} (h/L)$. Since $h/L <<1$ for our shallow layer flow, we can safely ignore the normal component of the velocity in our momentum balance equations. The $y-$ component of momentum balance equation then becomes
\begin{equation}
    v_x \frac{\partial v_y}{\partial x} + v_y \frac{\partial v_y}{\partial y} = - \frac{1}{\rho_b} \frac{\partial \sigma_{yy}}{\partial y} + g_y.
\end{equation}
Note that terms on LHS are $\mathcal{O} (v_{x0}^2 \frac{h}{L^2}) = \mathcal{O}(\frac{v_{x0}^2}{g h}) (\frac{h^2 g}{L^2}) \approx \mathcal{O}(Fr^2 (\frac{h}{L})^2 g)$. Since $h/L$ \textcolor{black}{used in this study $\sim \mathcal{O} (10^{-2})$}, the left hand side terms representing the convective acceleration can be safely ignored compared to gravitational acceleration in normal direction equation even if Froude number $Fr \approx \mathcal{O} (1)$. The $y-$ direction momentum balance equation then simplifies to $\frac{\partial \sigma_{yy}}{\partial y } + \rho_b g \cos\theta = 0$ since $g_y = -g \cos\theta$. Using free surface boundary condition at the free surface $h(x)$, we get $\sigma_{yy} = \rho_b g \cos\theta \int_y^{h(x)} dy = \rho_b g \cos\theta [h(x) - y]$. Since granular flows exhibit normal stress differences and accounting for their presence has been shown to improve the velocity predictions in case of periodic chute~\parencite{tripathi2011rheology,tripathi2013density}, we account for the fact that vertical normal stress $\sigma_{yy}$ is not equal to local pressure $P$ i.e., $\sigma_{yy} \neq P$. \textcite{tripathi2013density} have shown that pressure $P$ in unidirectional flows is related to vertical stress $\sigma_{yy}$ as $P = \sigma_{yy} (1 - a \tan\theta)$, where parameter $a$ accounts for the anisotropy of the stress in granular flows. Hence the $y-$ momentum balance equation reduces to 
\begin{equation}
    P(y) = \rho_b g \cos\theta (1 - a \tan\theta) [h(x) - y].
    \label{eq:pressure_monodisperse}
\end{equation}
With these assumptions, the simplified momentum balance equation in $x-$ direction can be written as 
\begin{equation}
    \rho_{b}v_x\frac{\partial v_x}{\partial x} = \rho_{b}g\sin{\theta} - \frac{\partial \tau_{yx}}{\partial y},
    \label{eq:vel}
\end{equation}
where $\tau_{yx}$ is the local shear stress and local pressure. Since the \textcolor{black}{ratio of the} flowing layer thickness ($h$) to the length of chute ($L$) is $\mathcal{O}(1/100)$, the pressure gradient in the flow ($x-$) direction $\frac{\partial P}{\partial x}\approx \frac{\rho_b g h}{L}$ is much smaller compared to gravitational acceleration in $x-$ direction. Hence we ignore pressure gradient term in equation~\ref{eq:vel}. \textcolor{black}{In the dense flow regime,} the shear stress ($\tau_{yx}$) and pressure ($P$ ) relate with each other through the effective friction coefficient, $\mu(I) = |\tau_{yx}|/P$. We utilize the JFP model proposed by~\cite{jop2006constitutivenature} to relate the effective friction coefficient $\mu(I)$ with the \textcolor{black}{dimensionless inertial number $I=\dot{\gamma}d_p/\sqrt{P/\rho_p}$} as follows:
\begin{equation}
    \mu(I) = \mu_s + \frac{(\mu_m-\mu_s) }{1+I_0/I},
    \label{eq:mu_I}
\end{equation}
where $\mu_s$, $\mu_m$, and $I_0$ are the rheological model parameters. \textcolor{black}{Inertial number $I$ essentially represents local shear rate $\dot\gamma$ non-dimensionalised by local pressure $P$ as well as particle size ($d$) and density ($\rho_p$). For the unidirectional flow considering in this study, $\dot \gamma = \frac{d v_x}{dy}$.} 
\textcolor{black}{We obtain the rheological parameters by fitting the nonlinear functional form given by the RHS of equation~\ref{eq:mu_I} to the $\tau/P$ vs $I$ data over a range of $\theta$. The obtained values of these parameters are close to those reported in \textcite{tripathi2011rheology}. Hence, we use $\mu_s = \tan(20.16^\circ)$, $\mu_m = \tan(37.65^\circ)$, and $I = 0.434$ as the rheological parameters in this study. Equations~\ref{eq:pressure_monodisperse} and~\ref{eq:vel} need to be solved with appropriate boundary conditions.} To predict the velocity field by solving equation~\ref{eq:vel}, the velocity profile obtained from DEM simulations at $x=0$ (see figure~\ref{fig:DEM_snaps_theta_29_rho_2.0}) is utilized as the inlet boundary condition. No-slip boundary condition is used at the chute base for all $x$ values, while a zero shear stress condition is applied at the free surface \textcolor{black}{i.e., at} $y=h(x)$. The corresponding boundary conditions are given by
\begin{equation}
   \begin{split}
      \text{Inlet}~~BC~1: v_{x}(0,y) = v_0(y),~~~\\
      \text{No slip}~~BC~2: v_{x}(x,0) = 0,~~~~\\
      \text{Free surface}~~BC~3: \tau_{yx}(x,h) =0.
   \end{split} 
    \label{eq:IC_BC_momentum}
\end{equation}
Using the boundary conditions given above  (equation~\ref{eq:IC_BC_momentum}), the momentum balance equation~\ref{eq:vel} coupled with the $\mu(I)$ rheology is solved using MATLAB \texttt{pdepe} solver to predict the velocity field. \textcolor{black}{Note that the shear stress gradient in equation~\ref{eq:vel} relates to the pressure given by equation~\ref{eq:pressure_monodisperse}, which requires the knowledge of layer height $h(x)$.} This layer height $h(x)$ at any location $x$ is obtained using the integral mass balance, i.e., $\int_0^{h(x)} \rho_b v_{x} dy = constant$. \textcolor{black}{The value of this constant is equal to the mass flow rate per unit width $\rho_b h(x) v_{x,avg}$ which is evaluated from the inlet velocity boundary condition. The average velocity is calculated as $v_{x,avg}(x) = (1/h(x))\int_0^{h(x)} v_{x} dy$}. Using this average velocity, we compute the chute length ($L_{flow}$) for fully developed flow as the length for which the average velocity reaches $95\%$ of the fully developed average velocity, i.e., $v_{x,avg}^{95} = 0.95  (v_{x,avg}^{\infty} - v_{x,avg}^{0}) + v_{x,avg}^{0}$.

\subsection{Segregation modeling using convection diffusion equation}
We also consider a binary mixture consisting of two species having particles of identical size $d$ and different densities flowing down a plane inclined at an inclination angle $\theta$. Following the previous studies~\parencite{grayandchugunov2006particle,fan2014modelling,xiao2016modelling,Duan2021}, we write the convection diffusion equation for the time evolution of the volume concentration $f_i$ of $i^{th}$ species as follows:
\begin{equation}
\frac{\partial f_i}{\partial t}
+ \nabla \cdot (\mathbf{v}_i f_i)
+\nabla\cdot\mathbf{J}^{diff}_i = 0
\label{eq:conv_seg_diff_1}
\end{equation}
where $\mathbf{v}_i = v_{i,x} \hat{\mathbf{x}} + v_{i,y} \hat{\mathbf{y}}$ is species velocity. $\mathbf{J}_i^{diff} = J_{ix}^{diff}\hat{\mathbf{x}} +  J_{iy}^{diff}\hat{\mathbf{y}}$ is the diffusion flux of species $i$ given as $J_{ix}^{diff} = -D \frac{\partial f_i}{\partial x}$ and $J_{iy}^{diff} = -D \frac{\partial f_i}{\partial y}$; where $D$ is diffusion coefficient. $v_{i,x} = v_x + v_{ix}^{seg}$ is the streamwise velocity of species $i$ (in the $x$-direction) with $v_x$ being the mean flow velocity and $v_{ix}^{seg}$ being the segregation velocity in the $x$-direction. For free surface flows, \textcolor{black}{the} segregation velocity in the flow direction is much smaller compared to the flow velocity due to convection i.e., $v_{ix}^{seg} \ll v_x $ and hence we assume $v_{i,x} = v_x$.
Similarly, $v_{i,y} = v_y + v_{iy}^{seg}$ is the velocity of species $i$ in the $y$-direction (normal to free surface) with $v_y$ being the mean flow velocity and $v_{iy}^{seg}$ being the segregation velocity in the $y$-direction. \textcolor{black}{Note that in order to remain consistent with our DEM results \textcolor{black}{that utilizes} a periodic boundary condition in $z$-direction}, \textcolor{black}{we ignore the gradients along the $z-$ direction}. By substituting these in equation~\ref{eq:conv_seg_diff_1}, we obtain  
\begin{align}
\frac{\partial f_i}{\partial t}
+ \frac{\partial}{\partial x}(v_x f_i)
+ \frac{\partial}{\partial y}(v_y f_i)
+ \frac{\partial}{\partial y}(v_{iy}^{seg} f_i) 
= D \frac{\partial^2 f_i}{\partial x^2}
+ D \frac{\partial^2 f_i}{\partial y^2}
\label{eq:seg_diff_i_simp}
\end{align}
Since the flow is \textcolor{black}{unidirectional}, there is no net transport in the vertical direction i.e., $v_y = 0$. \textcolor{black}{An order of magnitude estimate suggests that diffusion flux in $x-$ direction $J_{ix}^{diff} \sim \mathcal{O} (D/L^2) $ while that in $y-$ direction $J_{iy}^{diff} \sim \mathcal{O} (D/h^2) $ giving $J_{ix}^{diff} \sim \mathcal{O} [(h/L)^2 J_{iy}^{diff}]$. Hence for shallow flows considered in this study where $h/L \sim \mathcal{O} (10^{-2})$, $J_{ix}^{diff} << J_{iy}^{diff}$. This has also been shown by} \textcite{deng2018continuum} \textcolor{black}{and hence} the streamwise diffusion term $D \frac{\partial^2 f_i}{\partial x^2}$ can be neglected \textcolor{black}{in comparison to diffusion in vertical direction}. Following these, equation~\ref{eq:seg_diff_i_simp} simplifies as 
\begin{align}
\frac{\partial}{\partial x}(v_x f_i)
+ \frac{\partial}{\partial y}(v_{iy}^{seg} f_i) 
= D \frac{\partial^2 f_i}{\partial y^2}.
\label{eq:seg_diff_i_simp_1}
\end{align}
The \textcolor{black}{first} term in the left hand side expands as $v_x\frac{\partial f_i}{\partial x} + f_i\frac{\partial v_x}{\partial x}$. \textcolor{black}{Since $f_i \leq 1$ and $\frac{\partial v_x}{\partial x}$ varies little along $x$, its effect is very small compared to other terms of equation~\ref{eq:seg_diff_i_simp_1}.} Hence, we assume that $\frac{\partial}{\partial x}(v_x f_i) \approx v_x\frac{\partial f_i}{\partial x}$ and equation~\ref{eq:seg_diff_i_simp_1} reduces to
\begin{equation}
    v_x\frac{\partial f_i}{\partial x} + \frac{\partial}{\partial y}(v_{iy}^{seg} f_i) 
= D \frac{\partial^2 f_i}{\partial y^2}.
    \label{eq:conv_seg_diff_3}
\end{equation}
\textcolor{black}{Here, $v_{iy}^{\mathrm{seg}} f_i$ represents the segregation flux of the $i^{\mathrm{th}}$ species in the $y-$ direction, denoted by $J_{iy}^{\mathrm{seg}}$.} Following previous studies~\cite{tripathi2013density,Duan2021}, we \textcolor{black}{account for the variation of} diffusion coefficient with local shear rate ($\dot{\gamma}$) and local volume average diameter ($d_{mix}$) as $D = b\dot{\gamma}d_{mix}^2$. Here, $b$ is an empirical parameter that is obtained using DEM data and we use $b = 0.041$ as reported by \cite{tripathi2013density,sahu_kumawat_agrawal_tripathi_2023}. For \textcolor{black}{the present study utilizing} equal size and different density particle mixtures, $d_{mix} = d$. Following \cite{tripathi2013density,sahu_kumawat_agrawal_tripathi_2023,kumawat2025transient}, we use the particle-force-based approach which has been shown to successfully predict the steady as well as transient segregation in periodic chute flows. Using this approach, the segregation velocity expression for heavy species $v_{Hy}^{seg}$ in a binary mixture can be written as
\begin{equation}
    v_{Hy}^{seg} = -\frac{g\cos{\theta}}{c\pi\eta d}  (m_H-m_L)f_L,
    \label{eq:den_seg_v_seg}
\end{equation}
where $m_H$ and $m_L$ are the masses of the heavy and light species, respectively, and $f_L$ is the volume concentration of the light species in the mixture. The effective viscosity of the mixture $\eta$ is defined as the ratio of the shear stress $\tau_{yx}$ to the shear rate $\dot{\gamma}$, i.e., $\eta = \tau_{yx}/\dot{\gamma}$. The Stokes drag coefficient $c$ depends on the local solids fraction and is obtained by following~\cite{tripathi2013density}.
Substituting segregation velocity expression from equation~\ref{eq:den_seg_v_seg} along with the relation for diffusion coefficient $D = b \dot \gamma d^2$ in equation~\ref{eq:conv_seg_diff_3}, we obtain the following non linear partial differential equation for heavy species, 
\begin{equation}
    v_x\frac{\partial f_H}{\partial x} =  \frac{\partial}{\partial y} \left[ \frac{g\cos{\theta} }{c\pi\eta d} (m_H - m_L)f_L f_H + b \dot \gamma d^2 \frac{\partial f_H}{\partial y }
    \right].
    \label{eq:conv_seg_diff_den_seg}
\end{equation}
The concentration field of the heavy species $f_H$ is obtained by solving equation~\ref{eq:conv_seg_diff_den_seg} along with appropriate boundary conditions. The concentration of light species is then obtained as $f_L = 1 - f_H$. The required boundary conditions are given as 
\begin{equation}
\begin{split}
    \text{Inlet}~~BC~1:  f_{H}(0,y) = f_{H,in}(y);~~~~~~~~~~~~~\\
    \text{Zero flux}~~BC~2: J_{Hy}^{seg}(x,0) + J_{Hy}^{diff}(x,0) = 0;~~~~\\
   \text{Zero flux}~~BC~3:  J_{Hy}^{seg}(x,h) + J_{Hy}^{diff}(x,h) = 0.~~~~
    \end{split}
 \label{eq:conv_seg_diff_bc}
\end{equation}
Here, $h$ is the vertical height of the flowing layer that varies along the streamwise direction \textcolor{black}{i.e., $h = h(x)$}, \textcolor{black}{obtained using the integral mass balance of species, i.e., $\int_0^{h(x)} \rho_{mix}  v_{x} dy = constant$}. To determine the species concentration field, it is necessary to evaluate the local mixture velocity ($v_x$), local shear rate ($\dot \gamma$) and effective viscosity ($\eta$). These quantities are obtained by solving the momentum balance equations. The bulk density for binary density mixtures is given as $\rho_b = \phi \rho_{mix}$, where local volume average density $\rho_{mix} = \displaystyle \sum_{i=1}^N\rho_if_i $. \textcolor{black}{The bulk density of monodisperse grains $\rho_b$ remains constant, and the pressure is therefore calculated by equation~\ref{eq:pressure_monodisperse}. In contrast, the local bulk density of binary mixtures depends on the local mixture composition and varies spatially. Accordingly, the pressure is obtained as $ P(y) = g\cos{\theta}(1-a\tan{\theta})\int_y^{H(x)}\rho_{b}dy$.} For binary mixtures differing only in density, the effective friction coefficient depends on the generalized inertial number $I_{mix} = \dot{\gamma}d/\sqrt{P/\rho_{mix}}$ given as $\mu(I_{mix}) = \mu_s + \frac{(\mu_m-\mu_s) }{1+I_0/I_{mix}}$(\cite{tripathi2011rheology}). 

The species concentration and velocity fields are computed by solving the convection-segregation-diffusion equation (equation~\ref{eq:conv_seg_diff_den_seg}) and the momentum balance equation (equation~\ref{eq:vel}) simultaneously using the \texttt{pdepe} solver along with the inlet and boundary conditions given in equation~\ref{eq:conv_seg_diff_bc} and equation~\ref{eq:IC_BC_momentum}, respectively. The \texttt{pdepe} solver uses numerical discretization to approximate the derivatives by dividing the domain into finite grids in both vertical and streamwise directions. A total of $N = 200$ spatial grids along the $y-$ direction are used corresponding to a grid size of approximately one-twentieth of the particle diameter. We choose the grid size $\Delta x = 1d$ in the $x-$ direction. The detailed algorithm for solving these coupled equations using MATLAB \texttt{pdepe} solver is given in the supplementary material~\ref{app:Numerical_method}. \textcolor{black}{We compute the chute length required for fully developed segregation in binary mixtures, denoted by $L_{seg}$ as the streamwise distance at which the difference in average center of mass position of light and heavy species $(\Delta y_{com} = y_{com,L} - y_{com,H})$ attains $95\%$ of its saturated value, i.e., $\Delta y_{com}^{95} = 0.95 (\Delta y_{com}^{\infty} - \Delta y_{com}^{0}) + \Delta y_{com}^{0}$.}

\section{Results and discussion}
\label{sec:results}
\textcolor{black}{In this section, we present results of our continuum model predictions for different situations. We begin by discussing the results for monodisperse systems followed by those for a binary mixtures of different density particles. We also present a comparison between the predictions of the continuum model and discrete element method (DEM) simulations for these cases. We then use the continuum model to explore the effect of density ratio, mixture composition, and chute inclination. Finally we use these results to estimate the chute length required for \textcolor{black}{fully developed} segregation and discuss the importance of intercoupling of rheology and segregation.} \textcolor{black}{All the properties \textcolor{black}{reported in this section} are non-dimensionlised using particle size $d$ as the length scale, $\sqrt{d/g}$ as the time scale and $\sqrt{gd}$ as the velocity scale}. 
\subsection{Flow evolution of monodisperse mixtures}
\begin{figure}[h]
    \centering
   \includegraphics[scale=0.42]{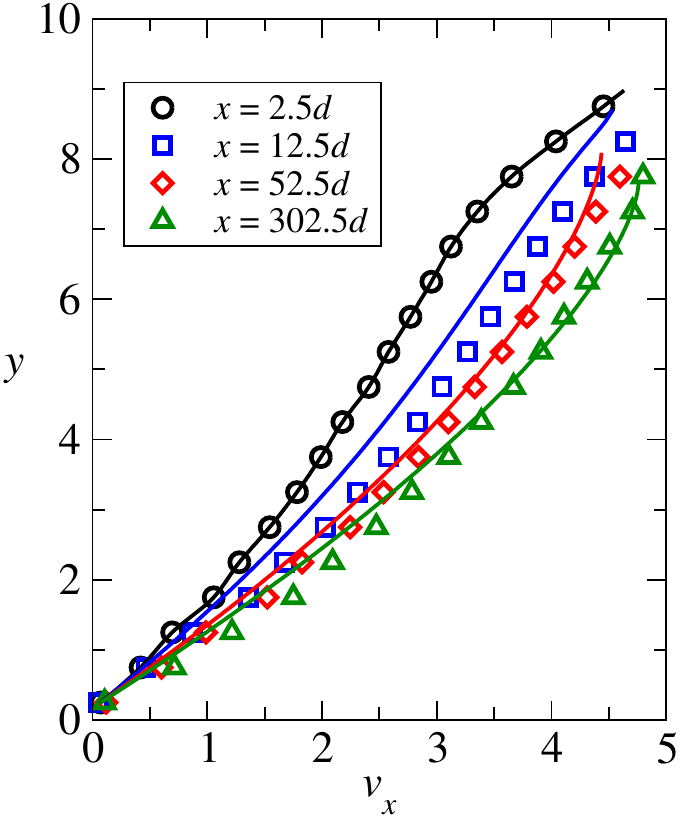}\put(-150,160){(a)}\quad \quad
     \includegraphics[scale=0.515]{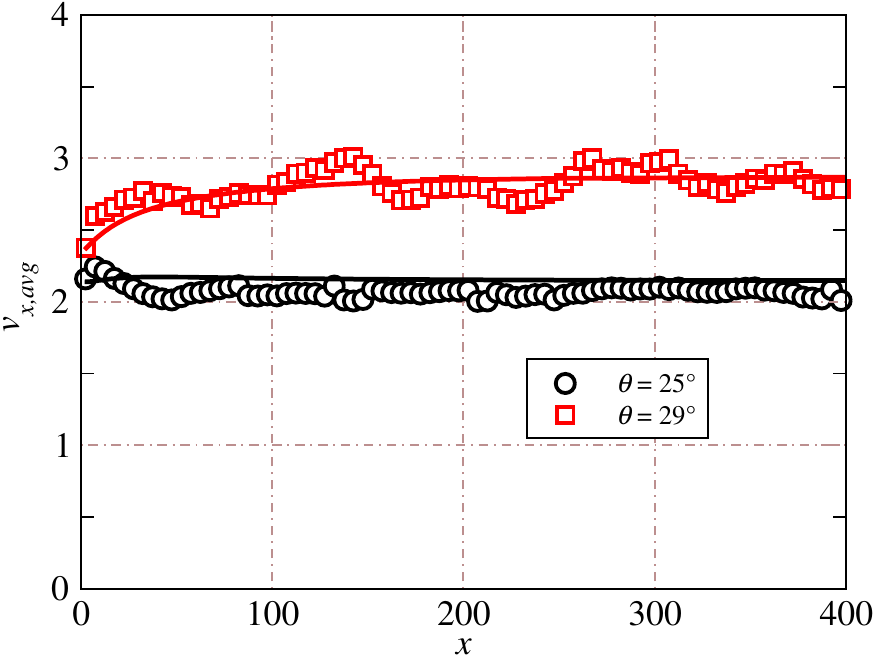}\put(-225,160){(b)} 
    \caption{(a) Steady-state velocity profiles at different chute locations for monodisperse grains flowing over an inclination angle $\theta = 29^\circ$. (b) Variation of average velocity ($v_{x,avg}$) with $x$ position along the chute. Symbols denote the DEM data, while solid lines represent the continuum model predictions.}
    \label{fig:rho_1.0_theta_25}
\end{figure}

\textcolor{black}{We first consider the flow of monodisperse grains over a long chute. The stream of particles falling from the generation volume \textcolor{black}{in our DEM simulations} gives rise to a thin layer of grains continuously flowing over the chute. After a short \textcolor{black}{duration}, the average kinetic energy of grains becomes nearly constant and we obtain a steady flow of grains along the chute.}
Figure~\ref{fig:rho_1.0_theta_25}a shows the steady-state velocity profiles at different chute locations for \textcolor{black}{this} monodisperse flow at \textcolor{black}{inclination angle }$\theta = 29^\circ$. 
Symbols represent the DEM results, and solid lines represent the continuum model predictions. \textcolor{black}{Black circles correspond to the velocity profile obtained from DEM simulations near the inlet (region $0 \leq x \leq 5d$). A smooth and continuous function fitted through this data is used as the inlet velocity boundary condition. The velocity profiles at different $x$ locations predicted from the model match well with the DEM data. As expected,} the velocity increases \textcolor{black}{with $x$, i.e., along the} downstream \textcolor{black}{direction} $x = 302.5d$. After that the velocity profiles remain nearly identical at different positions along the chute, \textcolor{black}{and the} height of flowing layer becomes \textcolor{black}{practically} constant. Figure~\ref{fig:rho_1.0_theta_25}b shows the variation of average velocity $v_{x,avg}$ along the chute. For $\theta = 29^\circ$, $v_{x,\mathrm{avg}}$ increases \textcolor{black}{upto same distance} and then attains an approximately constant value further downstream (red squares). The fluctuations in the DEM data are observed due to small undulations in the layer height, possibly due to the combined effect of discrete particle fluctuations \& surface waves \textcolor{black}{near the free surface} and the presence of baffles \textcolor{black}{near the base that are oriented} perpendicular to the flow direction. These fluctuations \textcolor{black}{in $v_{x,avg}$} get substantially reduced for $\theta = 25^\circ$ and the $v_{x,\mathrm{avg}}$ remains almost constant along the entire chute length at this angle. This is due to the fact that the inlet velocity profile obtained from DEM simulation is almost identical to the \textcolor{black}{steady-state} velocity profile that would be attained for $\theta = 25^\circ$ case. \textcolor{black}{Results presented in figure~\ref{fig:rho_1.0_theta_25} show that the continuum model is able to reliably predict the velocity evolution along the chute.}

Due to its much lower computational cost compared to DEM \textcolor{black}{simulations}, we utilize this continuum \textcolor{black}{model} to study the flow behaviour of granular materials under different conditions for long chutes.  
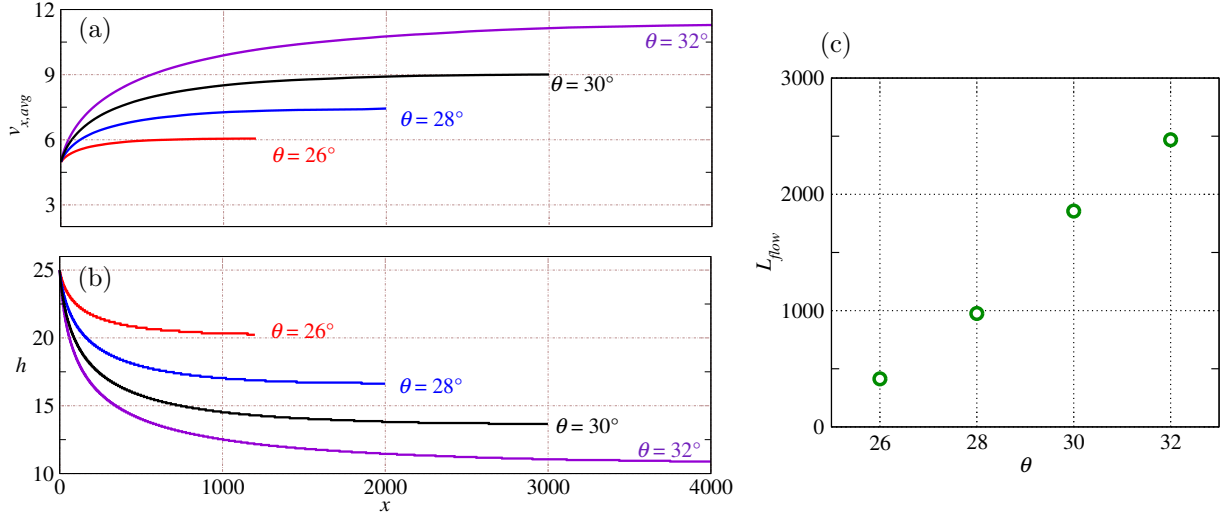
\begin{figure}[h]
    \centering
     \begin{tikzpicture}
        \node[anchor=south west] at (-3.08,3.75)
        {
            \includegraphics[scale=0.38]{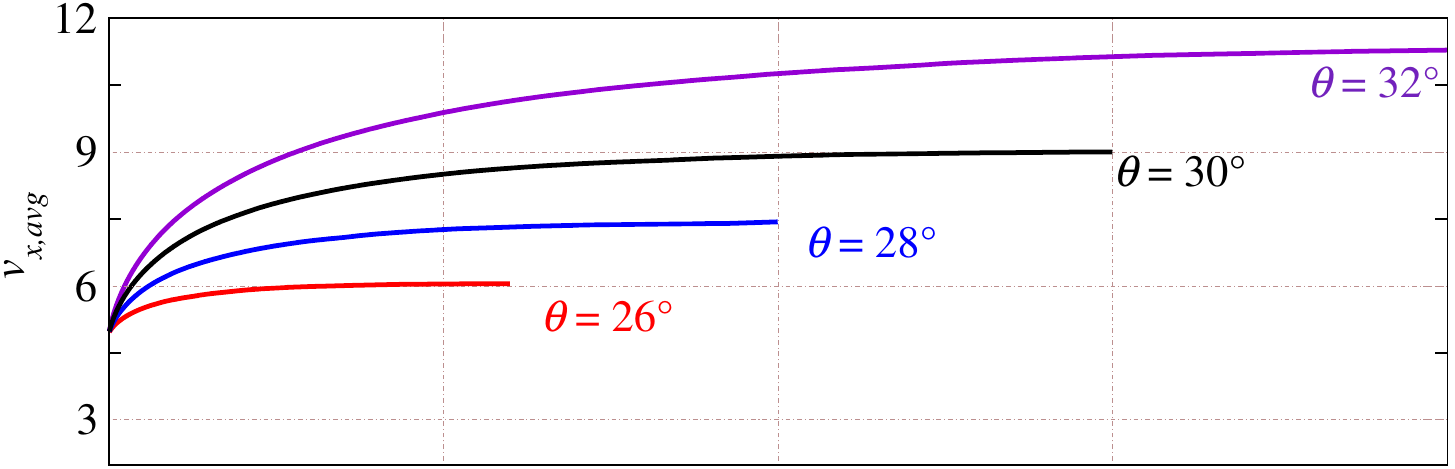}
        };
        \node at (-1.8,6.5) {(a)};
        \node[anchor=south west] at (-3.0,0)
        {
            \includegraphics[scale=0.38]{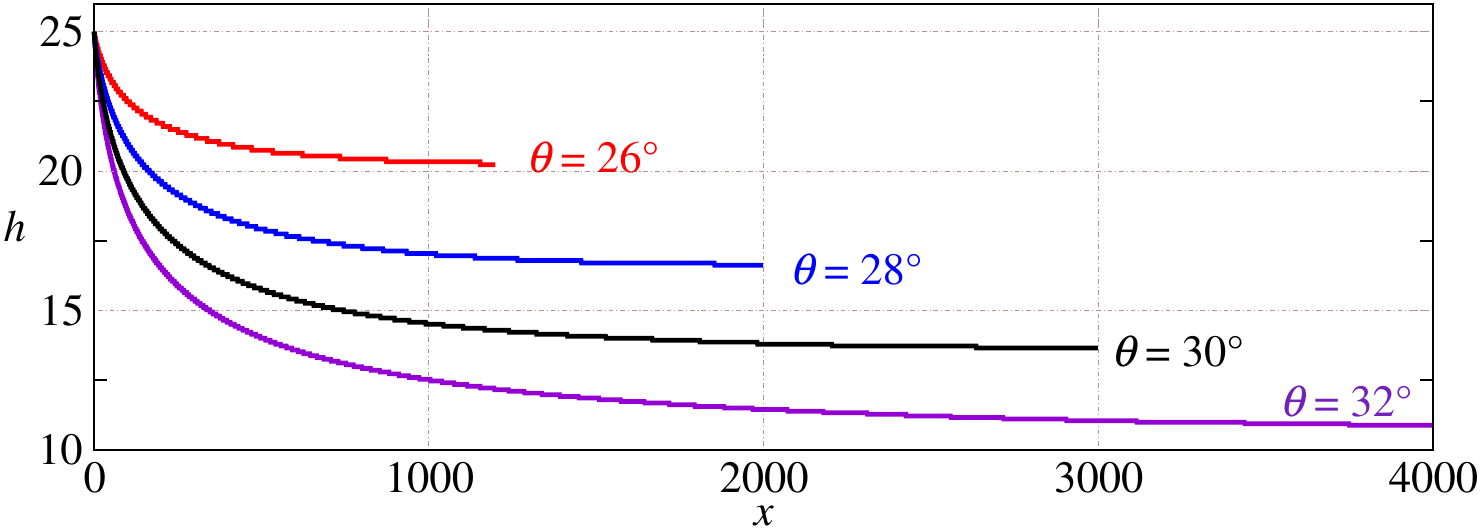}
        };
        \node at (-1.8,3.2) {(b)};
        \node[anchor=south west] at (6.8,0.5)
        {
            \includegraphics[scale=0.475]{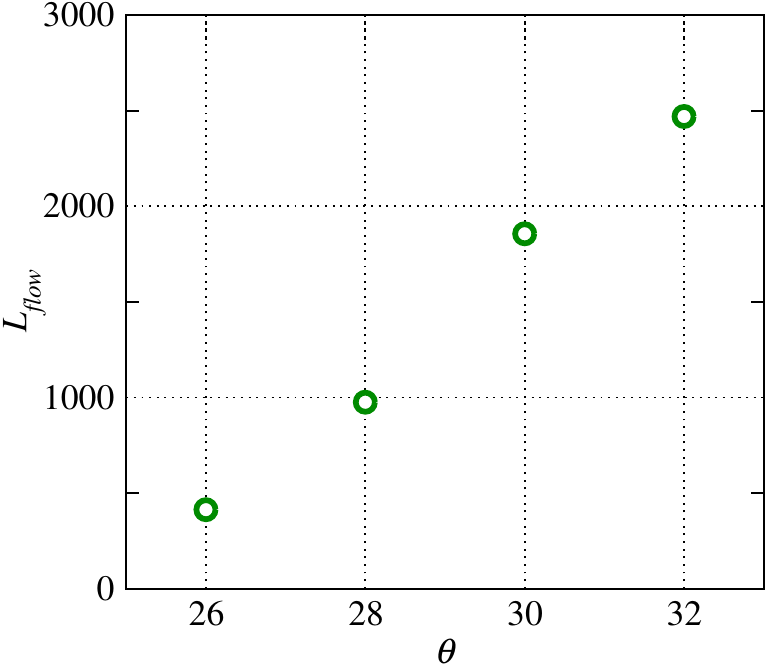}
        };
        \node at (8.0,6.3) {(c)};
    \end{tikzpicture}
    \caption{Evolution of the (a) average velocity $v_{x,avg}$ and (b) height of the flowing layer \textcolor{black}{along the chute} for different inclination angles \textcolor{black}{in the range} $\theta=26^\circ-32^\circ$ \textcolor{black}{predicted from} the continuum model. The inlet boundary conditions consist of a layer height of $h_i=25d$ and a linear velocity profile with an average velocity of $v_{x,avg}=5\sqrt{gd}$. (c) Corresponding variation of the fully developed chute length with inclination angle $\theta$.}
    \label{fig:long_chute_h_25d_v0_5}
\end{figure}
Rather than showing the velocity profiles at different locations along the chute, we \textcolor{black}{next} present the average velocity variation \textcolor{black}{along} the chute. Figure~\ref{fig:long_chute_h_25d_v0_5} shows the predictions of the continuum model for different inclination angles in the range $\theta=26^\circ - 32^\circ$. A fixed inlet flow rate having initial average velocity $v_{x,avg} = 5 \sqrt{g d}$ and flowing layer height $h_i =25d$ is used \textcolor{black}{for all these cases}. Figure~\ref{fig:long_chute_h_25d_v0_5}a and figure~\ref{fig:long_chute_h_25d_v0_5}b show the evolution of average velocity ($v_{x,avg}$) and layer height $h(x)$, respectively, along the chute. The average velocity $v_{x,avg}$ increases \textcolor{black}{and} the layer height $h$ decreases \textcolor{black}{with increasing the inclination angle}. Both $v_{x,avg}$ and $h$ gradually approach constant values along the chute length, indicating that the flow \textcolor{black}{eventually reaches} a fully developed state. The flow properties \textcolor{black}{show little variation} in the streamwise direction \textcolor{black}{beyond a length}. \textcolor{black}{The chute length required to obtain nearly fully developed flow, denoted as $L_{flow}$,} \textcolor{black}{corresponds to the length at which the average velocity increment reaches to $95\%$ of the maximum possible $v_{x,avg}$ at steady state.} The variation of this fully developed length $L_{flow}$ \textcolor{black}{(corresponding to data of figures~\ref{fig:long_chute_h_25d_v0_5}a and~\ref{fig:long_chute_h_25d_v0_5}b)} is shown in figure~\ref{fig:long_chute_h_25d_v0_5}c \textcolor{black}{for four different chute inclinations. Evidently,} $L_{flow}$ increases with increasing chute inclination, implying that steeper flows require a longer distance to attain a fully developed state due to stronger gravitational acceleration and \textcolor{black}{lower effective viscosity.}

\subsection{Evolution of segregation in binary density mixtures}
\label{subsec:Evolution_DEM_continuum}

\begin{figure}[h]
    \centering
       \includegraphics[scale=0.45]{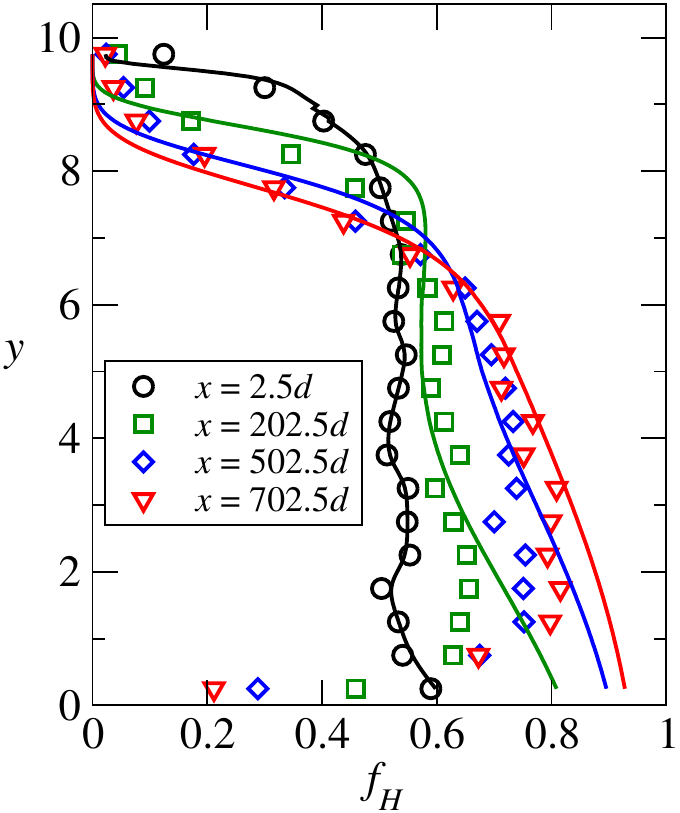}\put(-150,180){(a)}\quad \quad 
      \includegraphics[scale=0.45]{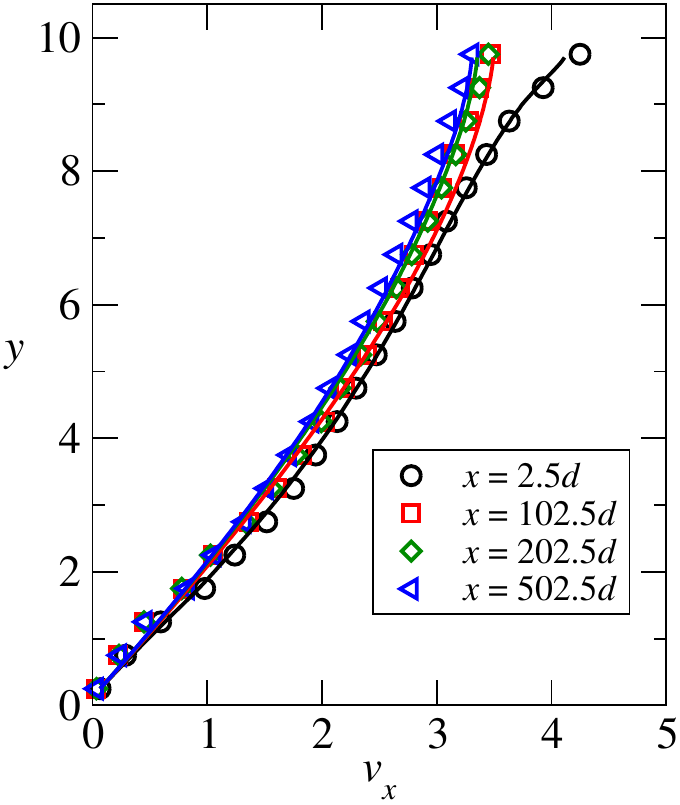}\put(-150,180){(b)}
    \caption{(a) Steady-state concentration profiles of heavy species and (b) Mixture velocity $v_x$ profiles in a binary mixture having $\approx51\%$ high-density particles with density ratio $\rho = 2.0$ at various chute locations for an inclination angle $\theta = 25^\circ$. Symbols denote the DEM data, while solid lines represent the continuum model predictions.}
    \label{fig:rho_2_theta_25}
\end{figure}

\textcolor{black}{We next report the results for binary mixtures differing in density with density ratio $\rho = 2.0$ flowing at an inclination angle of $\theta = 25^\circ$.} Figure~\ref{fig:rho_2_theta_25}a shows the  evolution of steady-state concentration profile of heavy species \textcolor{black}{and figure~\ref{fig:rho_2_theta_25}b shows that for} mixture velocity in \textcolor{black}{the flow} direction along the chute. Symbols denote the DEM data, while solid lines represent the continuum model predictions. Near the inlet (at $x = 2.5d$), the concentration of the heavy species remains nearly uniform, indicating an almost well mixed state of both species, as shown by the black symbols in figure~\ref{fig:rho_2_theta_25}a. \textcolor{black}{A smooth function fitted through this data (shown by black lines) is used as the inlet concentration profile in the continuum model to predict the evolution of concentration with $x$.}

Moving downstream along the chute, the symbols indicate that the concentration of high density particles decreases near the free surface and increases toward the base. The corresponding solid lines are in good agreement with the DEM data. \textcolor{black}{This confirms that} the particle-force-based continuum model accurately captures the evolution of concentration across the layer height \textcolor{black}{at different $x$ locations} along the chute.
Figure~\ref{fig:rho_2_theta_25}b shows the mixture velocity profiles at various \textcolor{black}{locations} along the chute \textcolor{black}{which} remains nearly unchanged \textcolor{black}{due to the inlet velocity being very close to the fully developed velocity profile at this inclination. Note that the continuum model is able to capture small decrease in the velocity at downstream locations observed in DEM simulations.} 

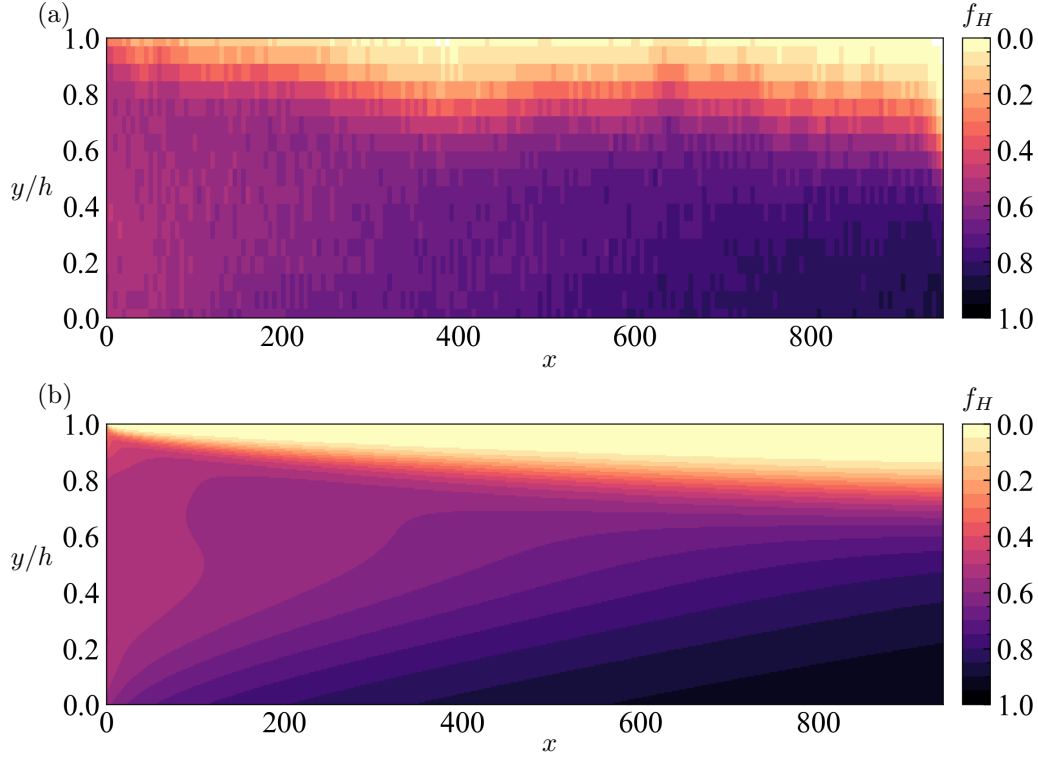
\begin{figure}[h]
    \centering
    \includegraphics[scale=0.7, trim=0 0 0 0, clip]{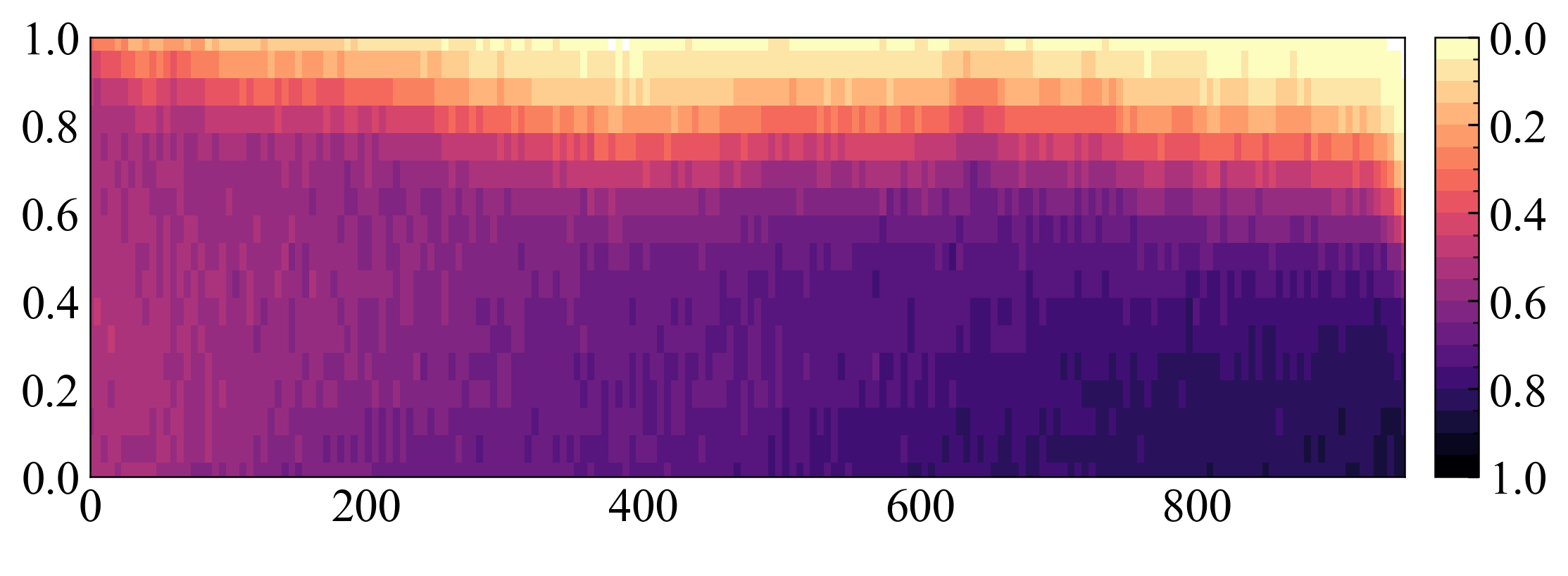}\put(-380,135){(a)}\put(-390,70){$y/h$}\put(-190,5){$x$}\put(-32,135){$f_H$}
      \\ 
     \includegraphics[scale=0.7, trim=0 0 0 0, clip]{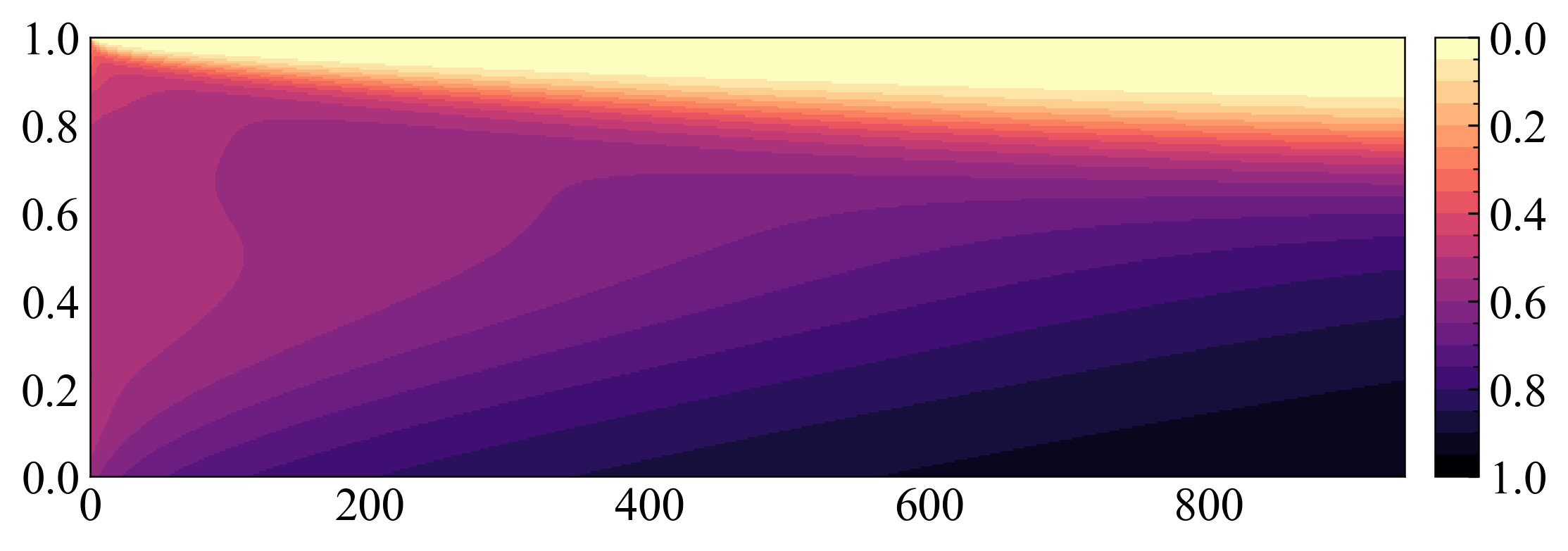}\put(-380,131){(b)}\put(-390,70){$y/h$}\put(-190,0){$x$}\put(-32,130){$f_H$}\\
    \caption{Color maps of the high-density species in a binary mixture having $\approx51\%$ high-density particles with density ratio $\rho = 2.0$, flowing down a chute inclined at $\theta = 25^\circ$ using (a) DEM simulation data, and (b) continuum model predictions.}
    \label{fig:colormaps_25}
\end{figure}

A comprehensive view of the evolution of species concentration along the entire chute \textcolor{black}{length} \textcolor{black}{can be obtained by plotting the concentration} color map of the high-density species. Figure~\ref{fig:colormaps_25}a shows \textcolor{black}{such a} color map of the high-density species concentration obtained from DEM data, while figure~\ref{fig:colormaps_25}b shows the corresponding color map for the continuum model predictions. \textcolor{black}{Near} the inlet \textcolor{black}{region}, the color distribution is \textcolor{black}{close to $0.5$} across the depth, indicating that the mixture \textcolor{black}{remains} in a \textcolor{black}{relatively well} mixed state \textcolor{black}{near the inlet}. As the flow progresses downstream, a clear separation of species is observed. The high-density species progressively migrates towards the bottom of the chute, as indicated by the darker shades near the base, while the upper region becomes rich in light-density particles, shown by the yellow color ($f_H \approx 0$).  The discrete nature of the color map in figure~\ref{fig:colormaps_25}a is due to the larger grid size \textcolor{black}{($\Delta y = 0.5d$)} used in DEM data \textcolor{black}{compared to that $\Delta y \approx 0.05d$ in figure~\ref{fig:colormaps_25}b}. 
\begin{figure}[htbp]
    \centering
    \includegraphics[scale=0.47]{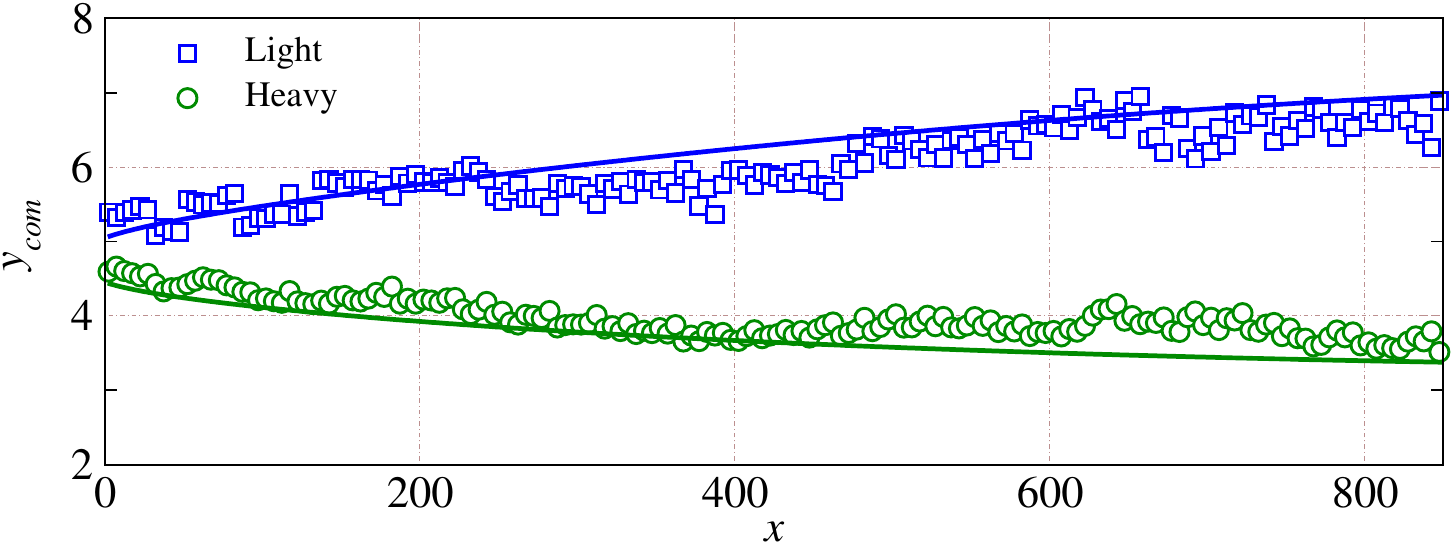}\put(-340,120){(a)}\hfill
     \includegraphics[scale=0.47]{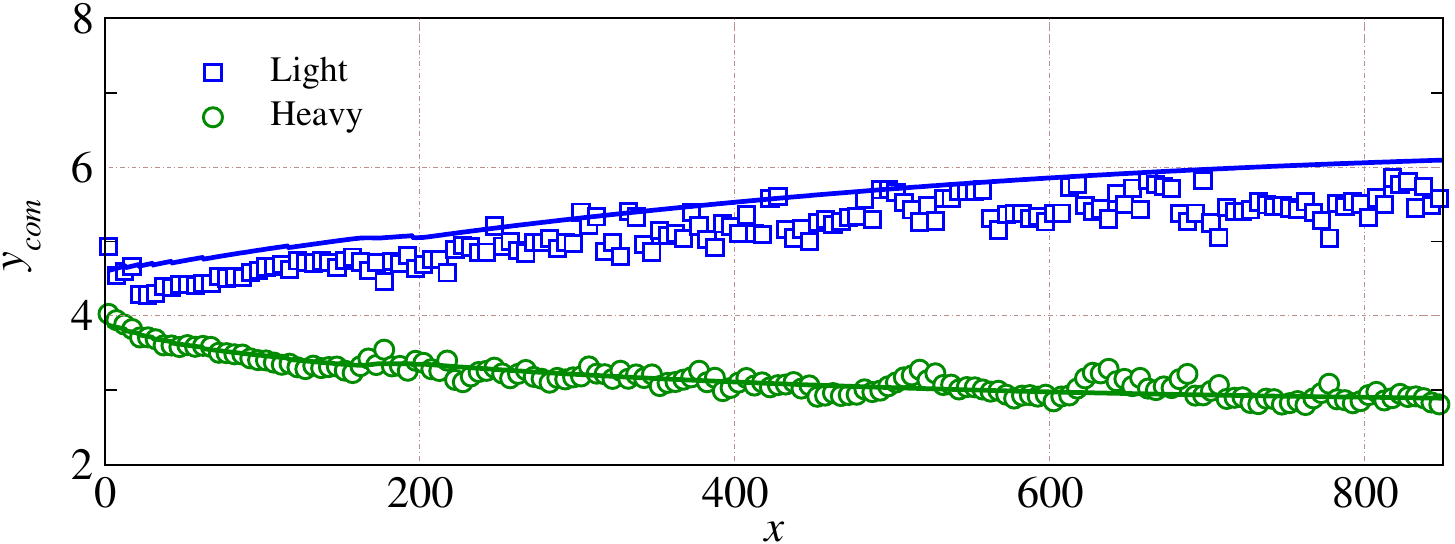}\put(-340,120){(b)}\hfill
    \caption{Evolution of species center of mass in binary mixture having $\approx51\%$ high-density particles \textcolor{black}{of} density ratio $\rho = 2.0$ at inclination angles (a) $\theta = 25^\circ$, (b) $\theta = 29^\circ$. Symbols represent the DEM data \textcolor{black}{and} solid lines show the continuum model predictions.}
    \label{fig:rho_2_theta_25_ycom}
\end{figure}
The variation of the center of mass of the species $y_{com}$ along the length of the chute for the same case is shown in figure~\ref{fig:rho_2_theta_25_ycom}. Figures~\ref{fig:rho_2_theta_25_ycom}a and~\ref{fig:rho_2_theta_25_ycom}b show the data for inclination angles $\theta = 25^\circ$ and $29^\circ$, respectively. Symbols represent the DEM data while solid lines correspond to the continuum model predictions. \textcolor{black}{Due to segregation, low density species concentrates near the free surface and} hence, $y_{com}$ increases \textcolor{black}{with $x$} for the low-density species. \textcolor{black}{In contrast, the $y_{com}$} of the high-density species \textcolor{black}{decreases with $x$ due to the settling of heavy grains near the base for} both inclination angles. The $y_{com}$ values at $\theta = 29^\circ$ are lower than those at $\theta = 25^\circ$, which \textcolor{black}{could be} attributed to the larger reduction in layer height \textcolor{black}{of the higher speed flow} at the \textcolor{black}{larger} inclination angle \textcolor{black}{resulting from the mass balance constraint}. As before, the continuum model predictions of $y_{com}$ match \textcolor{black}{well} with DEM data \textcolor{black}{for both angles}. Results shown in figures~\ref{fig:rho_2_theta_25}~-~\ref{fig:rho_2_theta_25_ycom} confirm that reliable predictions, \textcolor{black}{in good agreement DEM results, can be obtained from the continuum model}. Note that \textcolor{black}{a very} detailed comparison of these effects in periodic chutes using a similar approach has been reported in our previous works~\cite{tripathi2013density,kumawat2025transient} and good agreement between DEM and model predictions has been observed. \textcolor{black}{Hence, in this work, we avoid} performing computationally expensive DEM simulations for such large systems. \textcolor{black}{Instead,} we present only the continuum model results to \textcolor{black}{better} understand the effect of various parameters on segregation evolution \textcolor{black}{in the forthcoming sections}.

\subsection{Effect of density ratio, composition, and inclination angles}
\label{subsec:effect_density_comp_theta}
\begin{figure}[htbp]
    \centering
      \includegraphics[scale=0.62, trim=0 0 0 0, clip]{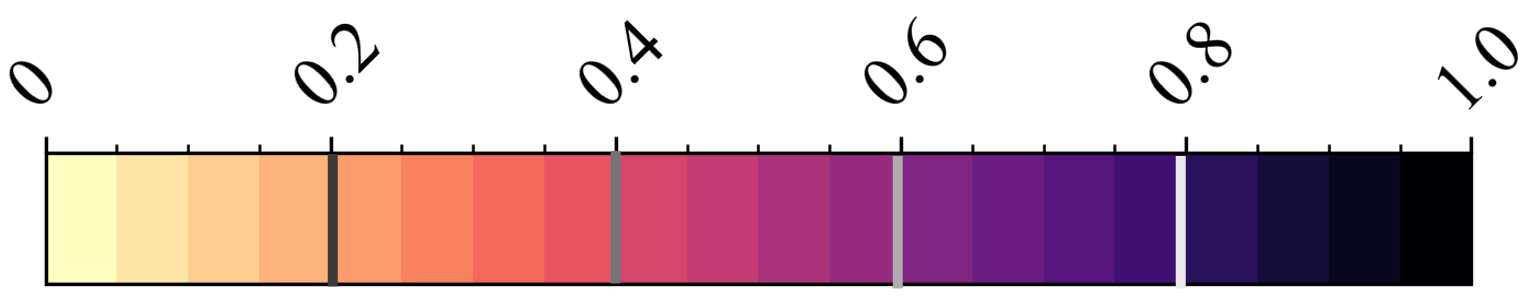}\put(-220,8){\large$f_H$}\\
       \includegraphics[scale=0.53, trim=0 0 0 0, clip]{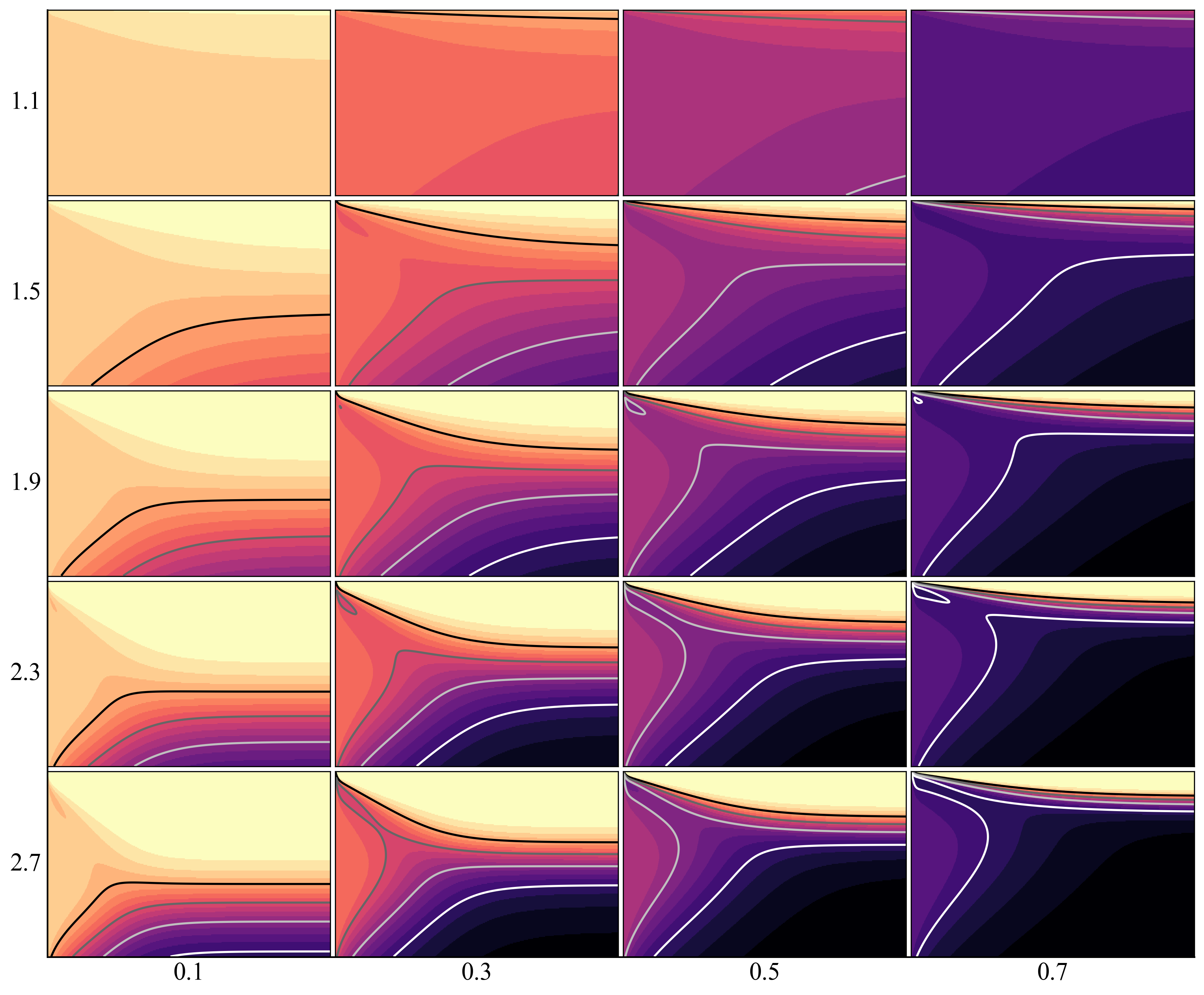}\put(-470,195){\large$\rho$}\put(-230,-5){\large$f_H^T$}
    \caption{Concentration color map of high density species for binary mixtures with density ratios in the range $1.1 \leq \rho \leq 2.7$ and total composition of high density particles $0.1 \leq f_H^T \leq 0.7$ at inclination angle $\theta = 29^\circ$. In each color map, the vertical axis spans $0 \leq y/h \leq 1$, while the horizontal axis spans $0 \leq x/d \leq 1000$.}
\label{fig:theta_29_rho_fl_effect}
\end{figure}

Figure~\ref{fig:theta_29_rho_fl_effect} shows the color maps of concentration of heavy species \textcolor{black}{for} binary mixtures flowing at inclination angle $\theta = 29^\circ$. In each color map, the vertical coordinate ranges from $0 \leq y/h \leq 1$, while the streamwise coordinate extends over $0 \leq x/d \leq 1000$. The density ratio is varied over the range $1.1 \leq \rho \leq 2.7$, and the total fraction of heavy species is varied within $0.1 \leq f_H^T \leq 0.7$. Each row corresponds to a particular density ratio $\rho$ which is reported at the left end of each row. Similarly, each column corresponds to a particular mixture composition which is reported at the \textcolor{black}{lower} end of each column. For all \textcolor{black}{the cases} considered, the same inlet velocity profile, obtained from the DEM data for binary mixture flowing at inclination angle $\theta = 29^\circ$, \textcolor{black}{is used}. A well-mixed state is considered at the inlet, i.e., $f_H = f_H^T$. This idealized well-mixed inlet differs from the DEM fitted profile used in section~\ref{subsec:Evolution_DEM_continuum} (figure~\ref{fig:rho_2_theta_25}a). This idealization is adopted to isolate the effects of density ratio, composition and inclination angle from residual inlet non-uniformity. \textcolor{black}{The black color contour lines correspond to $f_H = 0.2$ and dark grey color contour lines correspond to $f_H = 0.4$.} 
Similarly, light grey color contour lines correspond to $f_H = 0.6$ and white color contour lines correspond to $f_H = 0.8$. Top left panel in figure~\ref{fig:theta_29_rho_fl_effect} has no contour lines since $f_H < 0.2$ everywhere in this case (for $x \leq 1000d$). For \textcolor{black}{this case of} $f_H^T = 0.1$ and $\rho = 1.1$, segregation is barely noticeable even up to $x = 1000d$, indicating that a longer chute length is required for segregation to \textcolor{black}{fully}  develop. However, with increasing density ratio \textcolor{black}{(i.e., moving down column $1$)}, \textcolor{black}{different contour lines start appearing. This shows that} the degree of segregation becomes more pronounced along the chute length \textcolor{black}{with increase in $\rho$}, and for $\rho = 2.7$, complete segregation is clearly observed upto about $x = 600d$. \textcolor{black}{This enhanced segregation becomes evident with the appearance of darker shade regions with larger value of $f_H$ as we move downward along a other columns as well.} For all density ratios considered, the yellow region ($f_H \to 0$) decreases and darker region ($f_H \to 1$) increases \textcolor{black}{as we move along a row from left to right i.e.,} with increasing total composition of the heavy species, $f_H^T$. \textcolor{black}{At lower density ratios, the contour lines do not become flat at larger $x$ values indicating that the segregation is still evolving along the chute. At higher density ratios, the contour lines become horizontal towards the larger values of $x$, confirming fully developed segregation.}

\begin{figure}[H]
    \centering
      \includegraphics[scale=0.62, trim=0 0 0 0, clip]{Figs_paper/color_bar_small_labels_v3.png}\put(-220,8){\large$f_H$}\\
    \includegraphics[scale=0.53, trim=0 0 0 0, clip]{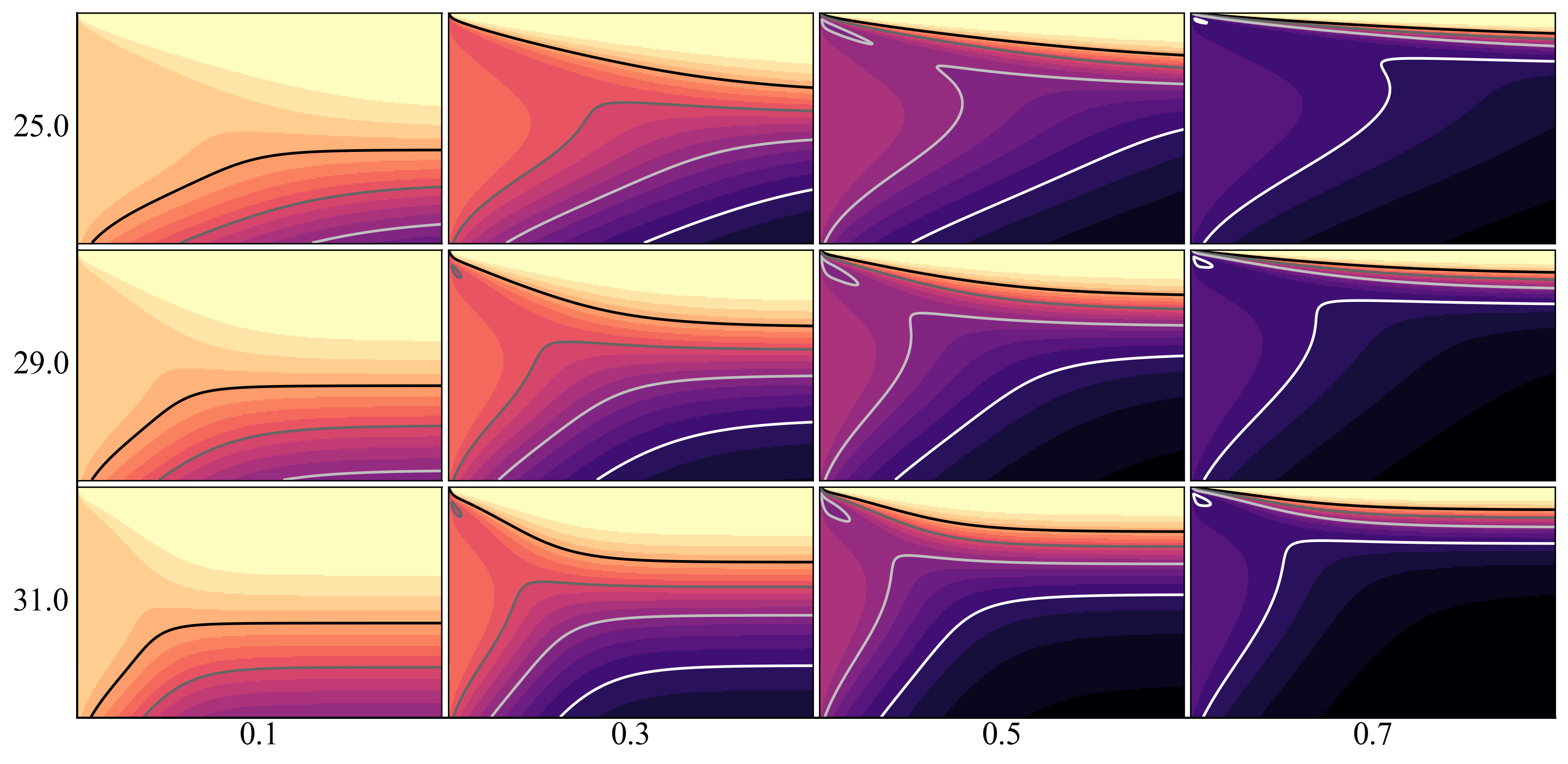}\put(-470,115){\large \textbf{$\theta$}}\put(-230,-10){\large$f_H^T$}
    \caption{Concentration color map of high density species for binary mixtures having density ratio $\rho = 2.0$ flowing at inclination angles in the range $25^\circ \leq \theta \leq 31^\circ$ and total composition of high density particles $0.1 \leq f_H^T \leq 0.7$. In each color map, the vertical axis spans $0 \leq y/h \leq 1$, while the horizontal axis spans $0 \leq x \leq 1000d$.}
    \label{fig:theta_fl_effect_rho_2}
\end{figure}

\textcolor{black}{We next report the effect of inclination angle on segregation evolution.} Figure~\ref{fig:theta_fl_effect_rho_2} shows the color maps of concentration of heavy species in a binary mixture having density ratio $\rho = 2.0$. In each color map, the vertical coordinate ranges from $0 \leq y/h \leq 1$, while the streamwise coordinate extends over $0 \leq x \leq 1000d$. The inclination angle is varied over the range $25^\circ \leq \theta \leq 31^\circ$, and the total fraction of heavy species is varied within $0.1 \leq f_H^T \leq 0.7$. For $\theta = 25^\circ$, the inlet velocity profile obtained from the DEM data at $\theta = 25^\circ$ is used, whereas for $\theta = 29^\circ$ and $31^\circ$, the inlet velocity profile obtained from the DEM data at $\theta = 29^\circ$ is used. Since the inlet \textcolor{black}{velocity} primarily sets the initial degree of development, we expect limited influence of this choice on the reported trends. For mixture rich in light-density particles ($f^T_H = 0.1$), \textcolor{black}{larger} yellow region ($f_H \to 0$) is present compared to mixture rich in high-density particles ($f_H^T = 0.7$). \textcolor{black}{As before, the increase in concentration of heavy particles leads to more darker shade regions as one moves along any row.} For a fixed chute length, the extent of segregation becomes more pronounced with increasing inclination angle.

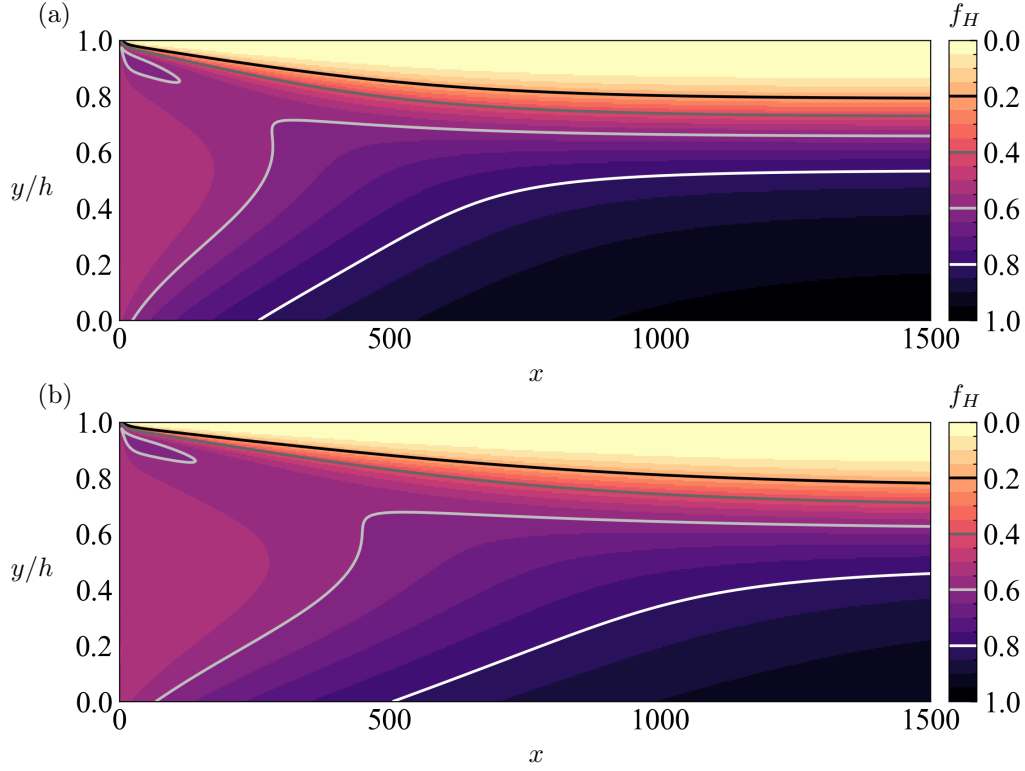
\begin{figure}[H]
    \centering
       \includegraphics[scale=0.70]{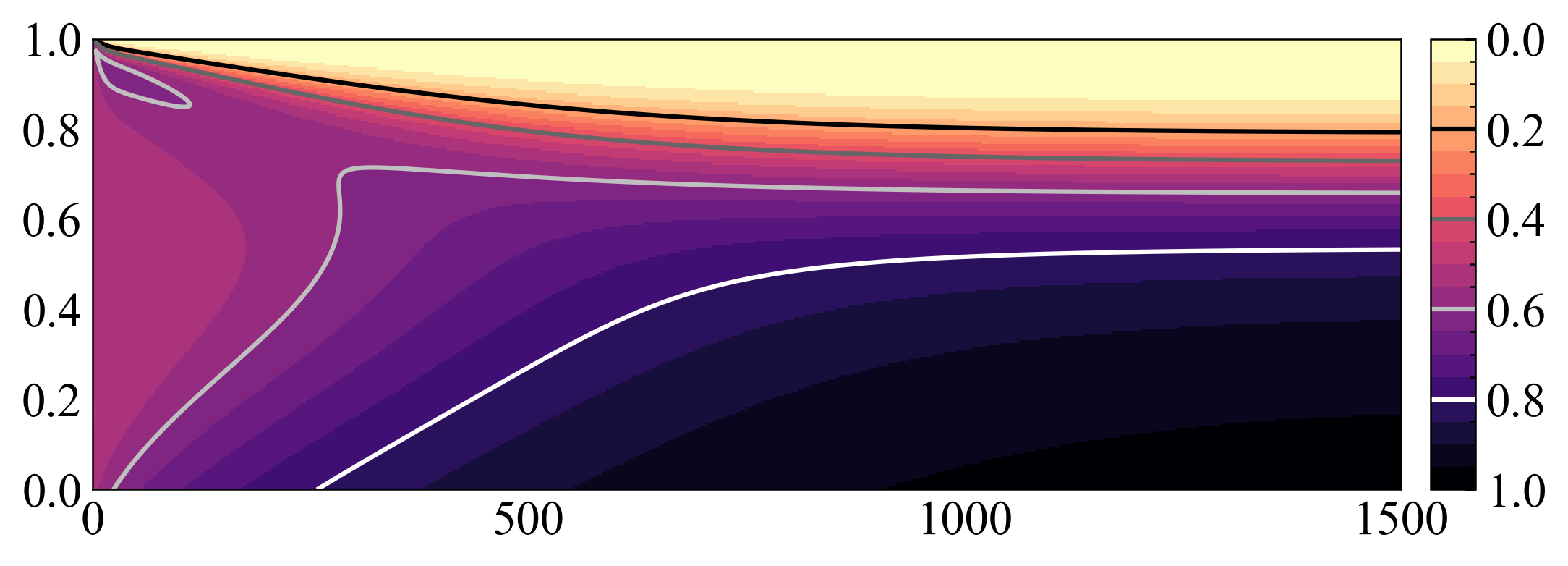}\put(-375,130){(a)}\put(-385,65){$y/h$}\put(-190,-5){$x$}\put(-32,130){$f_H$}\\
     \includegraphics[scale=0.70]{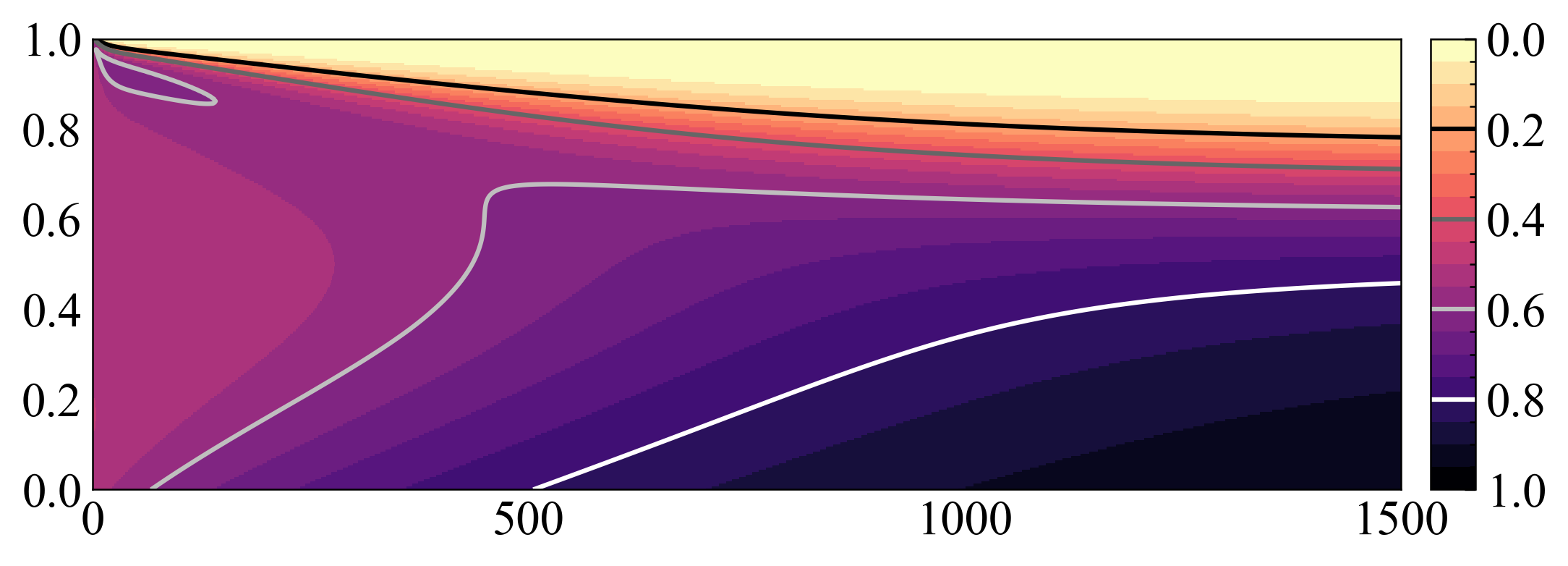}\put(-375,130){(b)}\put(-385,65){$y/h$}\put(-190,-5){$x$}\put(-32,130){$f_H$}
    \caption{Concentration color map of high density species for binary mixtures having density ratio $\rho = 2.0$ flowing at inclination angles $29^\circ$ for different basal slip conditions: (a) $5\%$ and (b) $30\%$ of the free surface velocity.}
\label{fig:theta_fl_effect_rho_2_slip}
\end{figure}

\textcolor{black}{We next explore the influence of basal slip on the extent of segregation using our continuum model. \textcolor{black}{The effect of basal roughness on segregation in different size mixtures has been investigated using DEM simulations~\cite{jing2017effect} as well as using a continuum model~\cite{deng2018continuum}.} Figure~\ref{fig:theta_fl_effect_rho_2_slip} shows the concentration color maps of high density species in equal composition binary mixture of density ratio $\rho = 2.0$ flowing at inclination angle of $\theta = 29^\circ$. 
Figures~\ref{fig:theta_fl_effect_rho_2_slip}a and~\ref{fig:theta_fl_effect_rho_2_slip}b show the continuum model predictions for basal slip velocities equal to $5\%$ and $30\%$ of the inlet free surface velocity observed in DEM simulations, respectively. The shear rate ($\dot{\gamma}$) across the layer is kept constant in both cases to isolate the influence of basal slip on segregation. The white contour corresponding to $f_H = 0.8$ appears at approximately $x \approx 250d$ for the $5\%$ slip case, whereas it emerges only around $x \approx 500d$ for the $30\%$ slip case. Likewise, the black-shaded region ($f_H \rightarrow 1$) appears \textcolor{black}{further downstream} with increasing basal slip, indicating slower segregation along the chute. A noticeable difference is also observed in the evolution of the light grey contour ($f_H = 0.6$), whereas the black ($f_H = 0.2$) and dark grey ($f_H = 0.4$) contours remain nearly identical between the two cases. Similar trends are observed for the other mixture compositions (not shown here). A larger chute length is required for the same mixtures to achieve the same degree of segregation at higher basal slip velocities. In other words, the extent of segregation can be reduced by increasing the basal slip. This trend is consistent with the size segregation study of \cite{zhou2016effect}, who reported that reducing the basal slip enhances segregation rate. Note that the authors found that this effect saturates when base friction coefficient exceeds interparticle coefficient. This saturation could possibly be attributed to the saturation of slip velocity at the base to its limiting value of zero slip velocity corresponding to no slip case.}

\begin{figure}[H]
    \centering
    \includegraphics[scale=0.455]{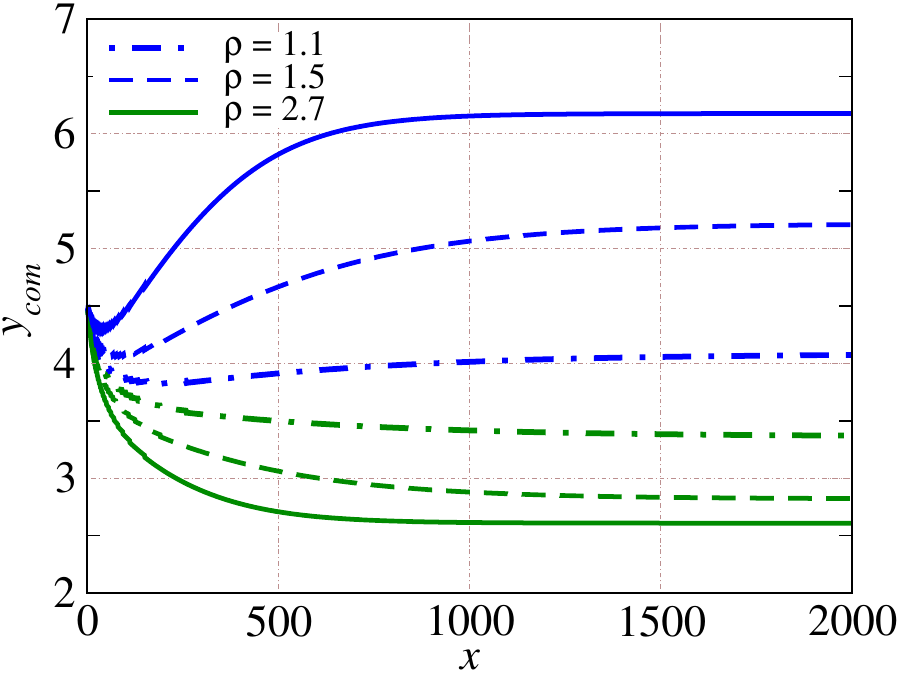}\put(-200,145){(a)}\put(-135,150){$f^T_H = 0.50; \theta = 29^\circ$} \quad
    \includegraphics[scale=0.455]{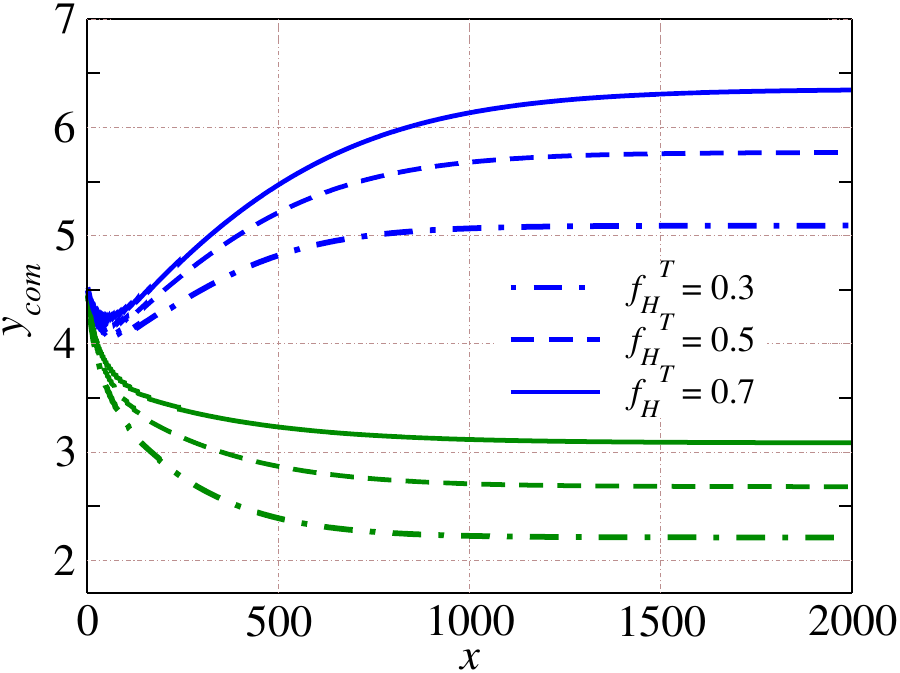}\put(-200,145){(b)}\put(-135,150){$\rho = 1.9; \theta = 29^\circ$}
    \\
     \includegraphics[scale=0.45]{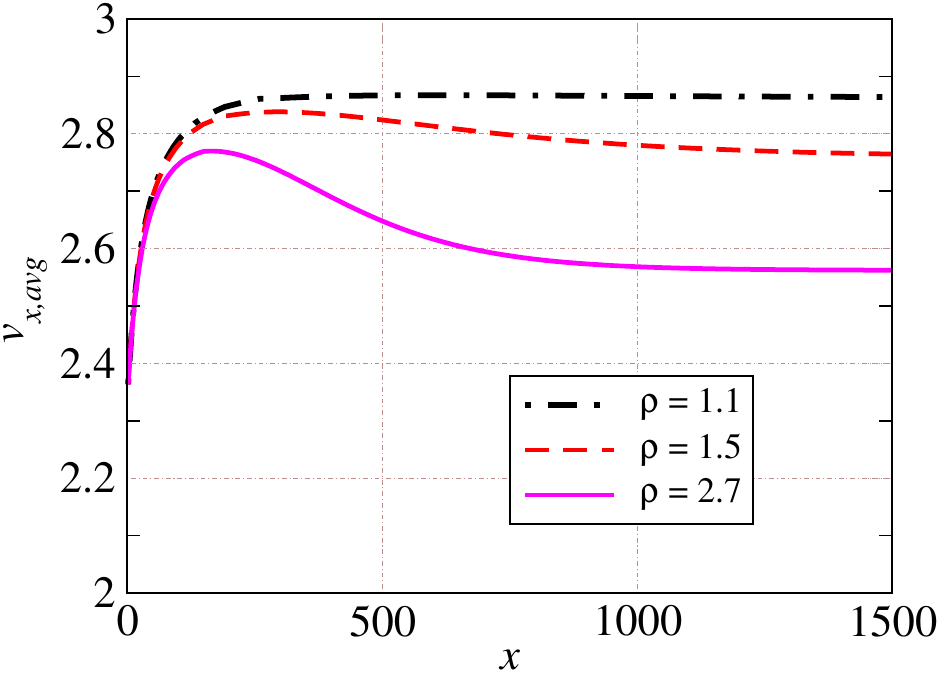}\put(-200,145){(c)}\quad
  \includegraphics[scale=0.45]{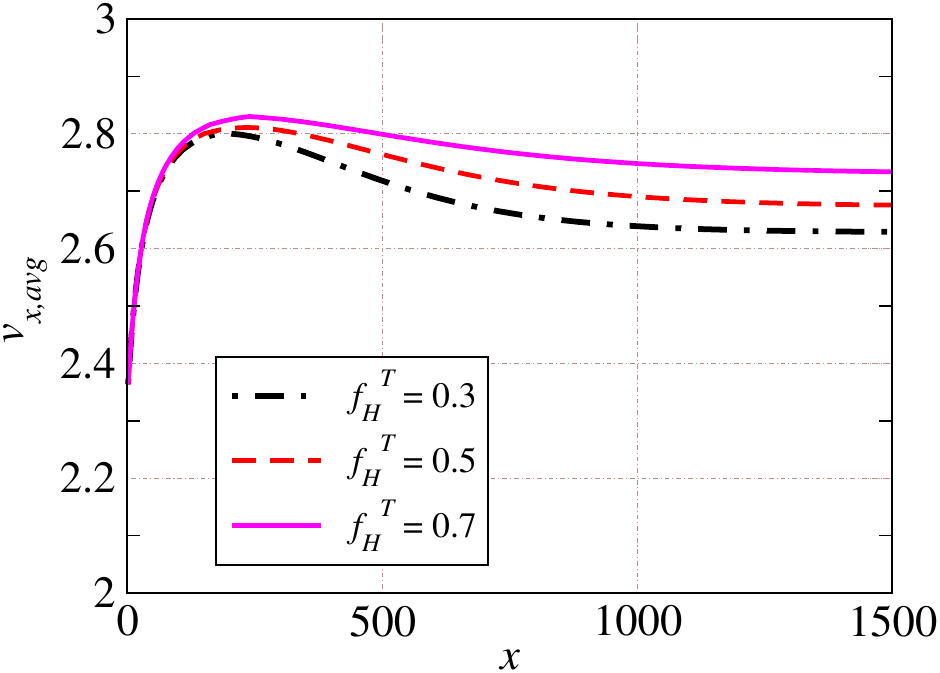}\put(-200,145){(d)}
\caption{Evolution of species center of mass $y_{com}$ \textcolor{black}{along with} chute length in equal composition binary mixture having (a) three different density ratios $\rho = 1.1$ (dashed-dotted lines), $\rho = 1.5$ (dashed lines), and $\rho = 2.7$ (solid lines) and (b) density ratio $\rho = 1.9$ for three different composition $f^T_H = 0.3$ (dashed-dotted lines), $f^T_H = 0.5$ (dashed lines), and $f^T_H = 0.7$ (solid lines). Green color represent the $y_{com}$ for heavy density species while blue color represent the $y_{com}$ for low density species \textcolor{black}{and results are reported for $\theta = 29^\circ$}. (c) and (d) represent the evolution of average mixture velocity \textcolor{black}{along the} chute length corresponding to (a) and (b), respectively.}
    \label{fig:ycom_vavg_theory}
\end{figure}

Figure~\ref{fig:ycom_vavg_theory} shows the evolution of the species center of mass $y_{com}$ and the average mixture velocity $v_{x,avg}$ along the chute length for selected cases corresponding to the continuum model predictions shown in figure~\ref{fig:theta_29_rho_fl_effect}. Figure~\ref{fig:ycom_vavg_theory}a shows the evolution of $y_{com}$ for an equal composition binary mixture with three density ratios, $\rho = 1.1$, $1.5$, and $2.7$, flowing down a chute inclined at $\theta = 29^\circ$. The green color represent the center of mass of the heavy species, whereas the blue color correspond to the light species. For $\rho = 1.1$ (dashed-dotted lines), the $y_{com}$ for both species evolves very slowly because of the weak density contrast. For $\rho = 1.5$ (dashed lines), the center of mass of the heavy species moves downward, while that of the light species shifts upward, and both attain constant values \textcolor{black}{around} $x \approx 2000d$. At density ratio $\rho = 2.7$ (solid lines), the segregation becomes significantly more pronounced, resulting in a larger separation between the centres of mass of the two species \textcolor{black}{that saturate around} chute length $x \approx 1000d$. Similarly, 
figure~\ref{fig:ycom_vavg_theory}b shows the evolution of species center of mass $y_{com}$ with chute length in binary mixture of density ratio $\rho = 1.9$ for \textcolor{black}{three} different mixture composition $f^T_H = 0.3, 0.5$, and $0.7$. At a given chute location, the $y_{com}$ values of both species are higher for mixtures rich in heavy particles than for mixtures rich in light particles. As the flow progresses downstream, the $y_{com}$ of the light species increases, whereas that of the heavy species decreases, and both eventually reach constant values, indicating a fully developed segregated state. \textcolor{black}{Further, mixtures with larger fraction of heavy species need larger distance for the segregation to evolve compared to mixture with lower concentration of heavy species.} Figures~\ref{fig:ycom_vavg_theory}c and~\ref{fig:ycom_vavg_theory}d show the evolution of the average mixture velocity $v_{x,avg}$ along the chute length corresponding to the cases shown in figures~\ref{fig:ycom_vavg_theory}a and~\ref{fig:ycom_vavg_theory}b, respectively. Figure~\ref{fig:ycom_vavg_theory}c shows the effect of density ratio on the flow development, whereas figure~\ref{fig:ycom_vavg_theory}d shows the influence of the mixture composition.

For $\rho = 2.7$ (solid lines in figure~\ref{fig:ycom_vavg_theory}c), the average velocity increases \textcolor{black}{along the} chute length, exhibits a local maximum, and then gradually \textcolor{black}{decreases to approach} a constant value. As the density ratio decreases, this local maximum becomes less pronounced. \textcolor{black}{In case of monodisperse grains, the peak in the average velocity vanishes as can be seen in figure~\ref{fig:long_chute_h_25d_v0_5}. This non-monotonic variation of the average velocity arises due to the intercoupling of segregation and rheology.} \textcite{kumawat2025transient} have shown that the average velocity during transient flow \textcolor{black}{in a periodic chute} initially increases, reaches a local maximum, and then gradually decreases to a steady-state value. \textcolor{black}{ The evolution of segregation during the flow affects} the layer viscosity, which further affects segregation evolution. A similar behaviour is observed here in the streamwise development of the flow.
A stronger influence of density contrast on the coupled behaviour of rheology and segregation \textcolor{black}{is observed} \textcolor{black}{compared to mixture composition}. \textcolor{black}{This is confirmed by figure~\ref{fig:ycom_vavg_theory}d which shows that} the average mixture velocity is only weakly affected by the overall composition of the mixture \textcolor{black}{since} only marginal variations are observed for different values of $f_H^T$.
\subsection{Comparison of segregation and flow evolution length scales}
\begin{figure}[H]
    \centering
    \includegraphics[scale=0.45]{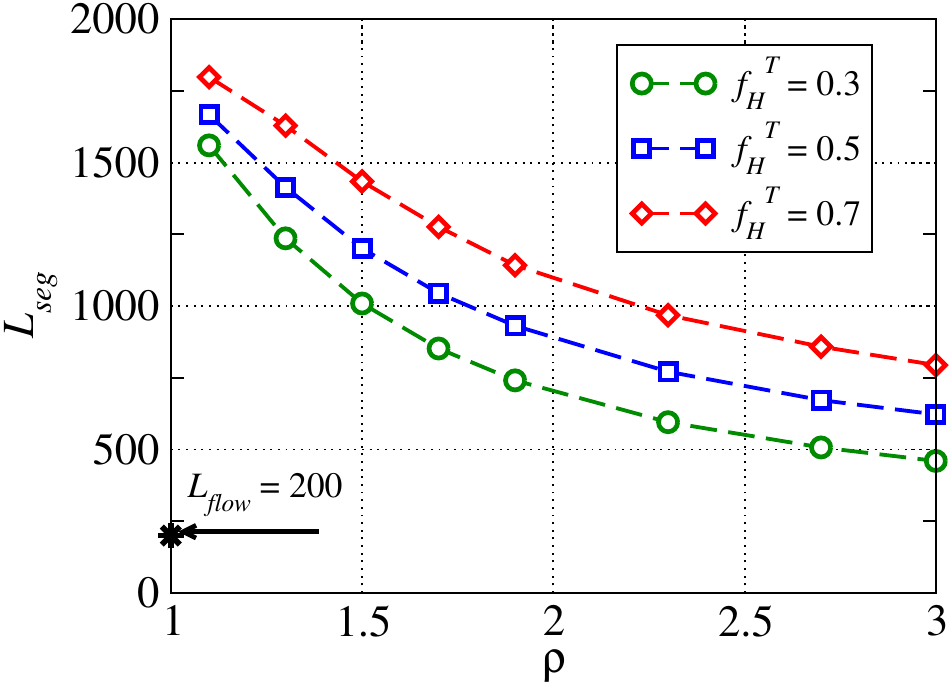}\put(-100,150){$\theta = 29^\circ$} \put(-190,155){(a)} \quad 
    \includegraphics[scale=0.45]{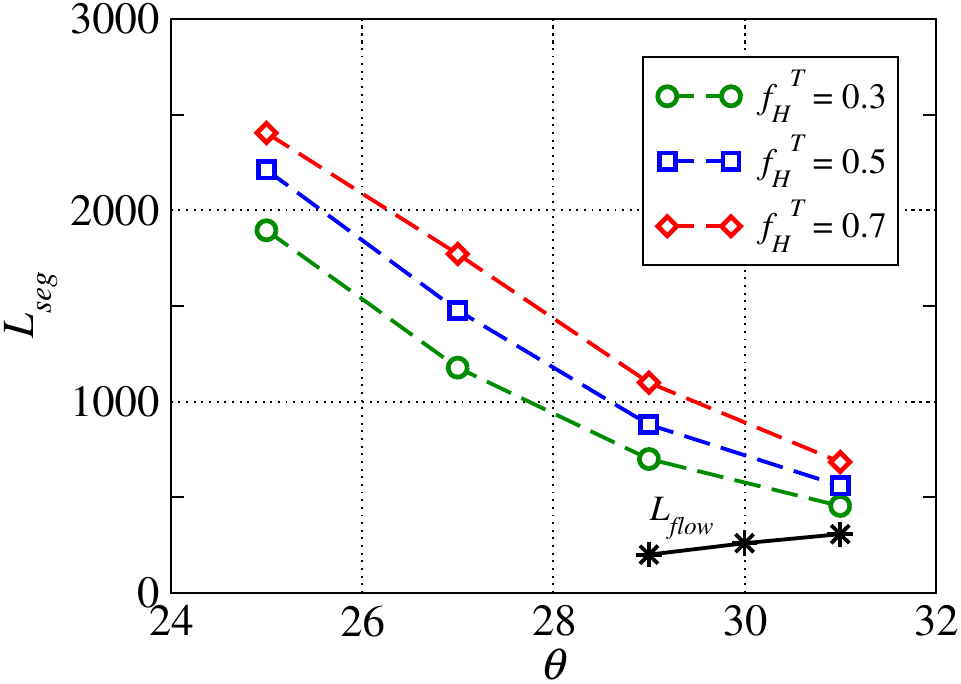}\put(-100,150){$\rho = 2.0$}\put(-190,155){(b)}
    \caption{Variation of chute length required for fully developed segregation with (a) density ratio for different \textcolor{black}{mixture} compositions at inclination angle $\theta = 29^\circ$, and (b) inclination angle for density ratio $\rho = 2.0$. Star symbols represent the chute length required for fully developed flow of monodisperse grains.}
    \label{fig:FDF_segregation}
\end{figure}

We next report the chute length required for fully developed segregation in figure~\ref{fig:FDF_segregation}.
Similar to the analysis of the chute length required for fully developed \textcolor{black}{flow} behaviour \textcolor{black}{of monodisperse grains}, we compute the chute length required to achieve fully developed segregation as the length at which \textcolor{black}{the change in} $\Delta y_{com}$ \textcolor{black}{with respect to its initial position} reaches $95\%$ of \textcolor{black}{its saturate value i.e., $\Delta y_{com}^{95} = 0.95 (\Delta y_{com}^{\infty} - \Delta y_{com}^{0}) + \Delta y_{com}^{0}$.}
Figure~\ref{fig:FDF_segregation} shows data for required chute length ($L_{seg}$) for fully developed segregation at different mixture compositions, $f_H^T = 0.3$, $0.5$, and $0.7$. 
Figure~\ref{fig:FDF_segregation}a presents the variation of $L_{seg}$ with density ratio at an inclination angle of $\theta = 29^\circ$, corresponding to data shown in figure~\ref{fig:theta_29_rho_fl_effect}. The length required for fully developed flow of monodisperse mixture ($\rho = 1.0$) is approximately $L_{flow} \approx 200$, shown by star symbol. This value is computed using the inlet velocity profile obtained from DEM simulations at $\theta = 29^\circ$ (figure~\ref{fig:rho_1.0_theta_25}a), ensuring a consistent basis for comparison with density ratio shown in figure~\ref{fig:theta_29_rho_fl_effect}. \textcolor{black}{Note that the value of $L_{flow}$ is significantly lower than that reported in figure~\ref{fig:long_chute_h_25d_v0_5}c due to different inlet conditions.}
For low density ratios, the chute length required for segregation to fully develop is much larger than that required for rheological development. 
However, for high density ratios, the length scale associated with segregation development becomes comparable to that required for rheology. A similar trend is observed for all mixture compositions considered. Thus, unlike empirical approaches~\cite{xiao2016modelling,deng2018continuum,Duan2021} and kinetic theory-based approaches~\cite{larcher2013segregation,larcher2015evolution}, it becomes essential to account for rheological evolution of the flow by solving momentum balance equations along with the convection-diffusion-segregation equation to accurately predict the evolution of flow and segregation. Similarly, figure~\ref{fig:FDF_segregation}b shows the variation of $L_{seg}$ with inclination angle $\theta$ at a density ratio of $\rho = 2.0$, corresponding to the data shown in figure~\ref{fig:theta_29_rho_fl_effect}. As the inclination angle increases, the chute length required for segregation to become fully developed decreases. At $\theta = 31^\circ$, the value of $L_{flow}$ is comparable to the $L_{seg}$, indicating the importance of \textcolor{black}{accounting} for the rheological evolution of the flow for accurate prediction of segregation evolution. 

\begin{figure}[bhtp]
    \centering
       \includegraphics[scale=0.70]{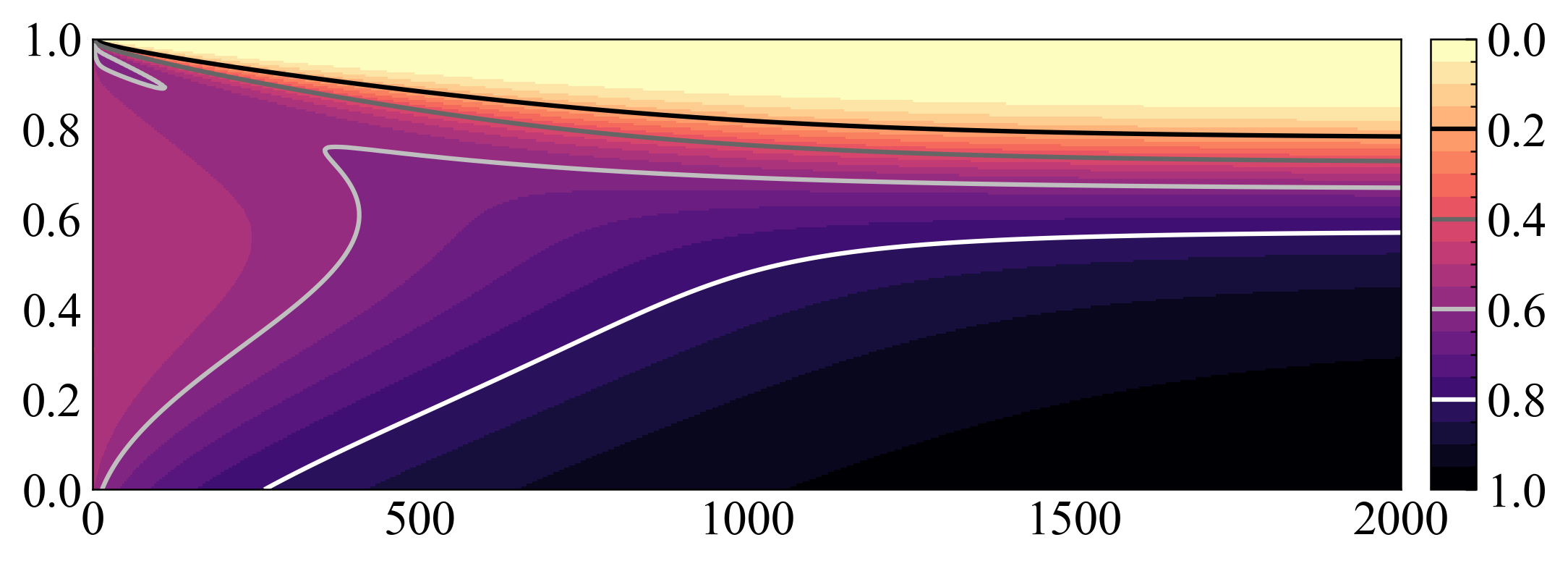}\put(-375,130){(a)}\put(-385,65){$y/h$}\put(-190,-5){$x$}\put(-32,130){$f_H$}\\
        \includegraphics[scale=0.70]{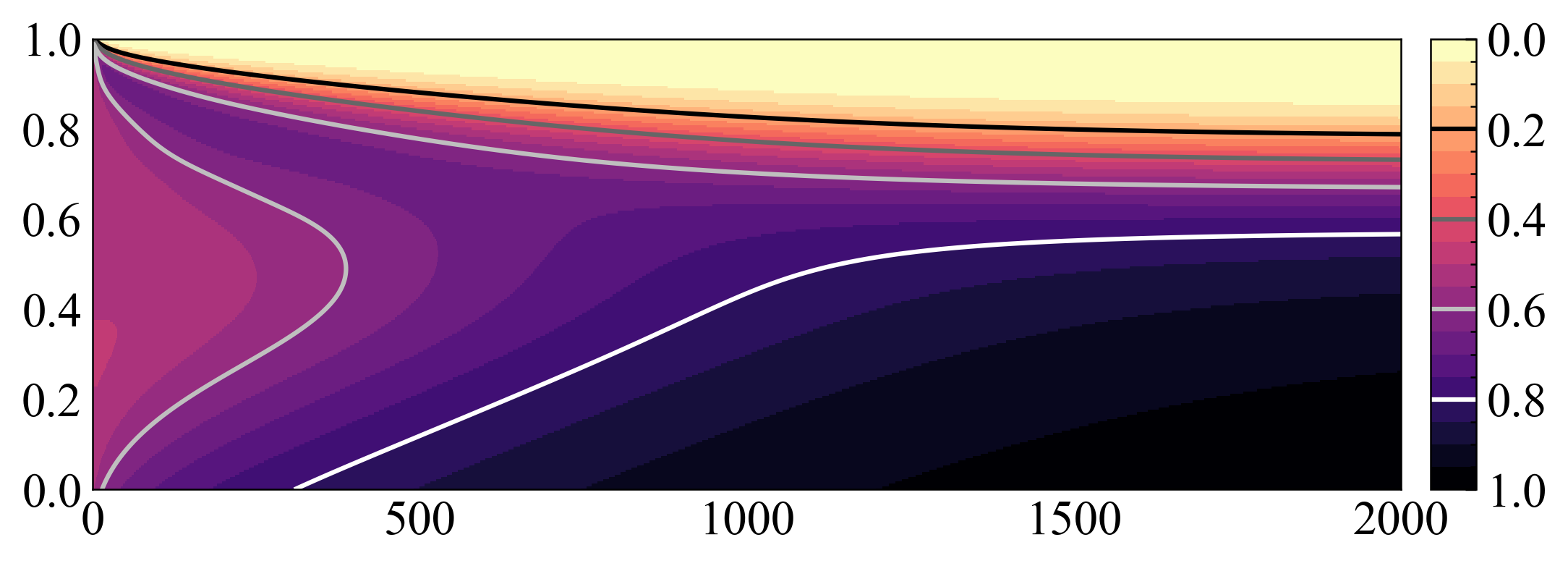}\put(-375,130){(b)}\put(-385,65){$y/h$}\put(-190,-5){$x$}\put(-32,130){$f_H$}
    \caption{Concentration color maps of the high-density species in a binary mixture having $50\%$ high-density particles with density ratio $\rho = 2.0$, flowing down a chute inclined at $\theta = 25^\circ$ (a) with and (b) without intercoupling of segregation and rheology.}
\label{fig:with_without_evolution_theta25}
\end{figure}
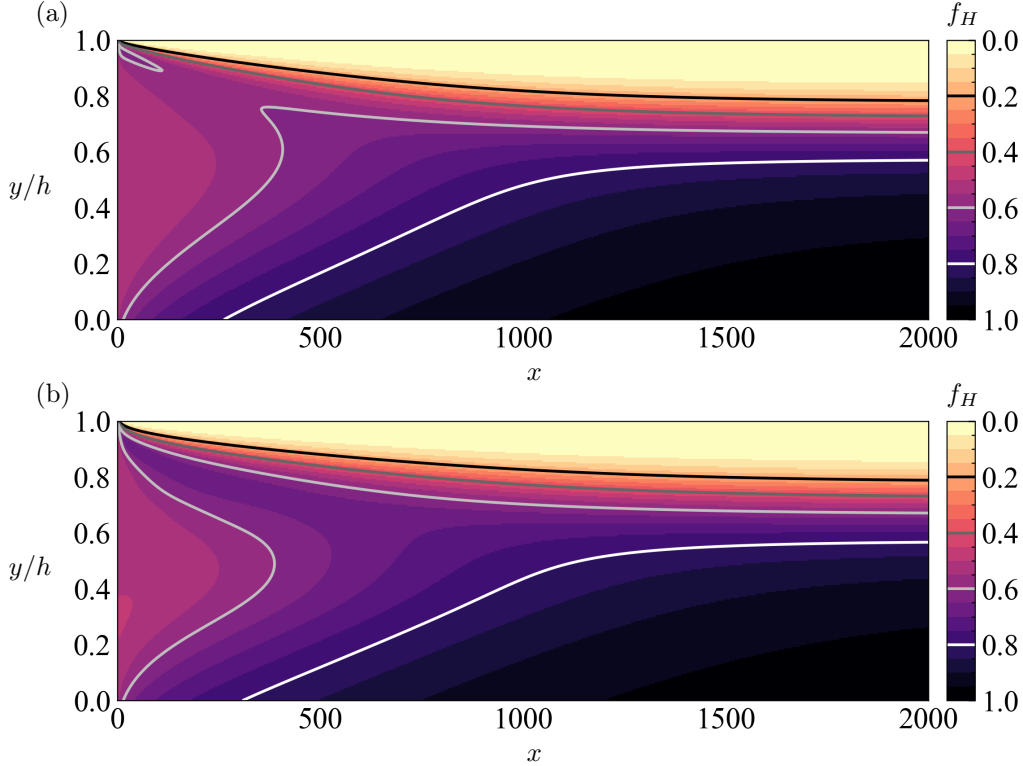

\textcolor{black}{To demonstrate the significance of incorporating the spatial evolution of the velocity field, we compare the continuum model predictions obtained using the evolving velocity field with those based on the assumption of a fully developed velocity field. Figure~\ref{fig:with_without_evolution_theta25} shows the concentration color map for heavy species in equal composition mixture of density ratio $\rho = 2.0$ flowing at inclination angle $\theta = 25^\circ$. The contour lines for $f_H = 0.2, 0.4, 0.6,$ and $0.8$ are denoted by black, dark grey, light grey, and white \textcolor{black}{color lines}, respectively. Since the flow development length is much smaller than the segregation development length at this inclination, the continuum model predictions obtained using the evolving velocity field (figure~\ref{fig:with_without_evolution_theta25}a) are almost identical to those obtained using a fully developed velocity field (figure~\ref{fig:with_without_evolution_theta25}b). 
The contour lines corresponding to $f_H = 0.2, 0.4, 0.6,$ and $0.8$ show minor differences in the entrance region ($x<500$). Beyond this region, the predictions obtained using the evolving and fully developed velocity fields are nearly identical. Hence, the evolution of velocity field can be safely neglected for the cases where $L_{flow} << L_{seg}$.} 

\begin{figure}[H]
    \centering
       \includegraphics[scale=0.70]{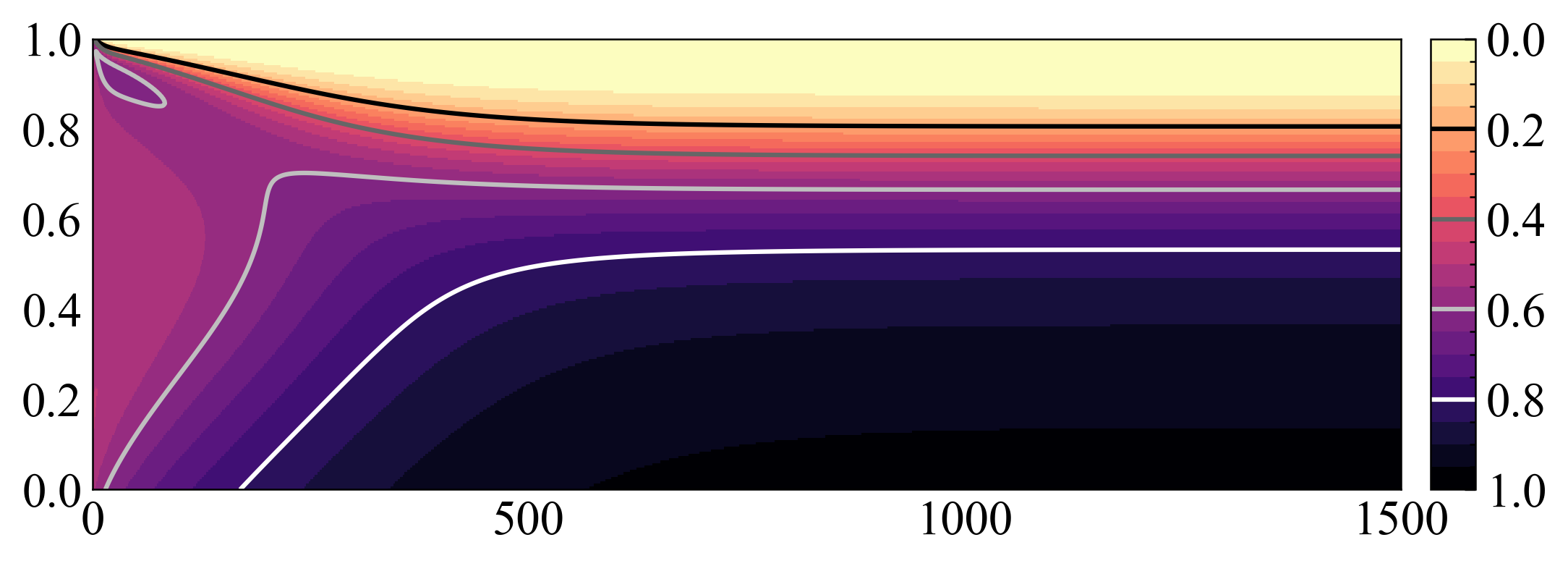}\put(-375,130){(a)}\put(-385,65){$y/h$}\put(-190,-5){$x$}\put(-32,130){$f_H$}\\
        \includegraphics[scale=0.70]{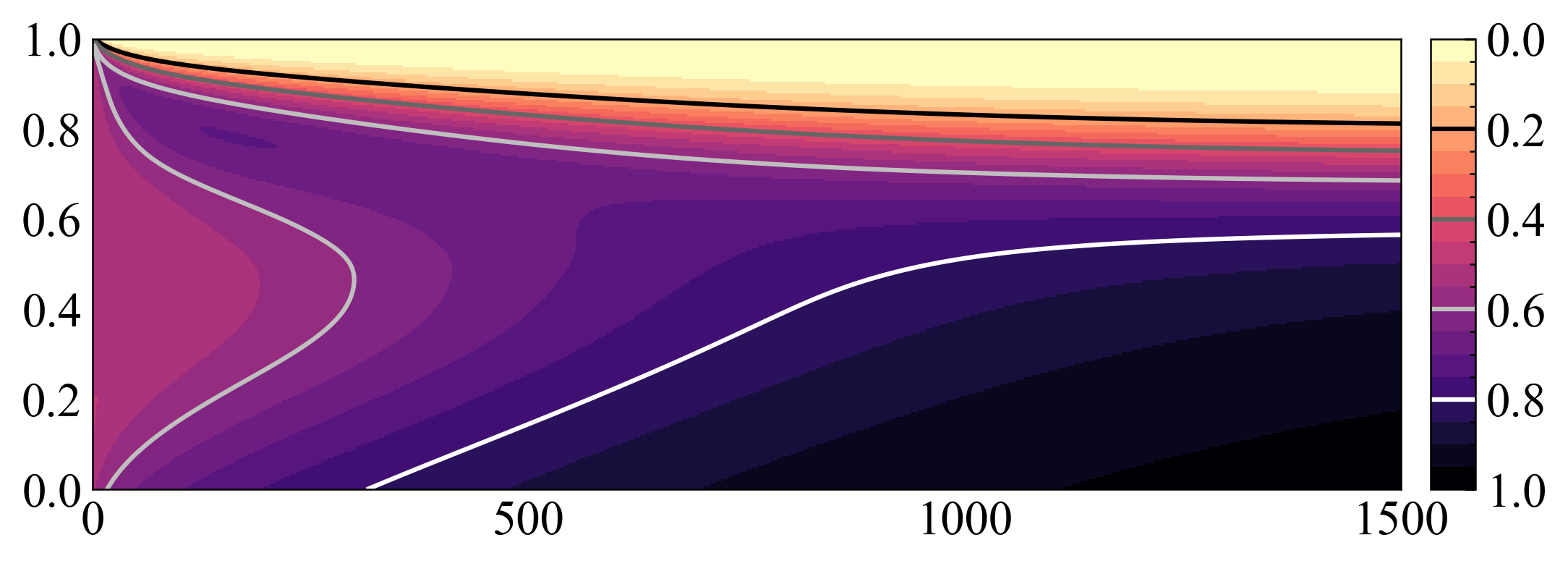}\put(-375,130){(b)}\put(-385,65){$y/h$}\put(-190,-5){$x$}\put(-32,130){$f_H$}
    \caption{Concentration color maps of the high-density species in a binary mixture having $50\%$ high-density particles with density ratio $\rho = 2.0$, flowing down a chute inclined at $\theta = 31^\circ$ (a) with and (b) without intercoupling of segregation and rheology.}
    \label{fig:with_without_evolution}
\end{figure}
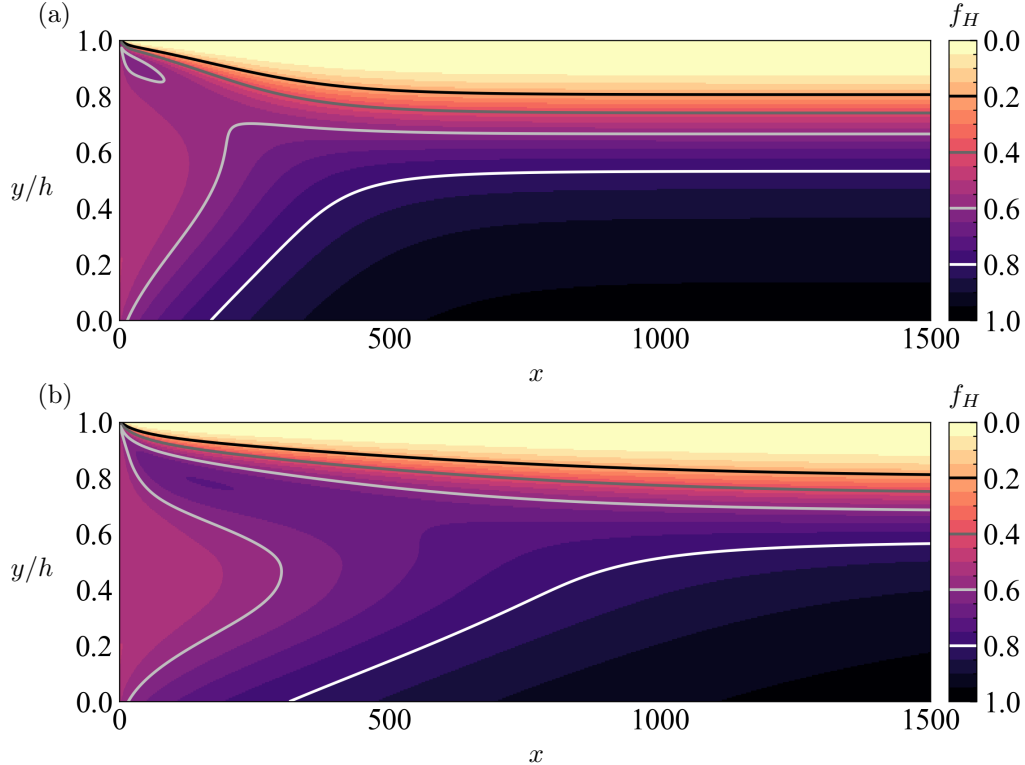

\textcolor{black}{Figure~\ref{fig:with_without_evolution} shows the spatial evolution of the concentration of the heavy species in an equal composition binary mixture with a density ratio of $\rho = 2.0$ flowing down an inclined plane at $\theta = 31^\circ$. Figure~\ref{fig:with_without_evolution}a shows the continuum predictions obtained by accounting for the streamwise evolution of the velocity field, whereas figure~\ref{fig:with_without_evolution}b shows the predictions obtained by assuming a fully developed velocity field throughout the chute. As expected, noticeable differences between the two are observable along the chute length since $L_{flow}$ becomes comparable to $L_{seg}$ \textcolor{black}{in this case}. The segregation evolves slowly along the chute when the continuum model predictions are obtained using the assumption of a fully developed velocity field. 
A similar trend has been reported by~\textcite{kumawat2025PandGtransient} who predicted the transient segregation of binary size mixtures in a periodic chute. Their continuum model predictions using the steady-state velocity field exhibited a noticeably slower segregation evolution than the DEM results for larger size ratios ($r = 1.5~\&~2.0$). 
A slower temporal evolution of segregation compared with DEM simulations have also been observed by 
\textcite{kumawat2025transient} when the authors utilized the kinetic-theory-based segregation model proposed by \textcite{larcher2015evolution} along with a fully developed velocity profile. These observations suggest that neglecting the evolution of the flow field can significantly affect segregation predictions when the characteristic flow development length is comparable to the segregation development length.}

\section{Conclusions}
\label{sec:conlusions}
In this work, we investigated the rheological and segregation behaviour of monodisperse as well as binary density mixtures flowing over an inclined plane. We developed a continuum model to solve steady-state momentum balance equations along with the $\mu-I$ rheological model~\cite{jop2006constitutivenature} to predict the flow over a long chute without sidewalls.
To check the validity of the continuum model, a few DEM simulations were also performed, and the model predictions were found to be in good agreement with the DEM results. For monodisperse grains, the continuum model successfully captured the streamwise evolution of the velocity profile, indicating that the rheological behaviour is accurately incorporated in the continuum framework. Using the $\mu(I)$ rheology parameters, the model was able to predict the evolution of the average velocity and layer height for different situations. We \textcolor{black}{could also estimate} the chute length required for the flow to attain a fully developed flow state. 

We next incorporated the particle-force-based segregation model into the continuum framework of solving the steady-state convection-diffusion-segregation transport equation along with the momentum balance equations and investigate the density segregation of binary granular mixtures flowing over a long chute. For binary density mixtures, the continuum framework was able to describe the coupled evolution of flow and segregation along the chute. \textcolor{black}{The model predicts} the effect of density ratio, mixture composition, inclination angle and basal slip on segregation behaviour. Consistent with the findings of \textcolor{black}{previous works}~\cite{jing2017effect,deng2018continuum}, \textcolor{black}{our} continuum model predicts a reduction in segregation with increasing basal slip. At low density ratios (\textcolor{black}{and/or} low inclination angles), segregation develops gradually and requires a larger chute length, whereas at higher density ratios (\textcolor{black}{and/or} higher inclination angles) segregation becomes more pronounced and the corresponding fully developed segregation length decreases. Therefore, in case of high density and/or high inclination angles, it becomes important to account for the evolving rheology by solving the momentum balance equations together with the convection-diffusion-segregation equation. The \textcolor{black}{popular approach used in literature that assumes fully developed flow throughout the chute while predicting segregation may lead to substantial errors in such cases}. 

\textcolor{black}{A straightforward extension of this work is to employ the particle-force-based size segregation model and} investigate the flow and segregation of different size mixtures over a long chute by accounting for the coupled role of segregation and rheology. In future studies, this continuum framework can be extended to predict the behavior of rapid granular flows, which are computationally expensive to simulate using DEM, by incorporating $\mu(I)$ rheological model as proposed by~\textcite{patro2021}. An important limitation of the current study is its neglect of the presence of sidewalls. Hence, the chute lengths required for fully developed flow and segregation reported in this work correspond to \textcolor{black}{no sidewalls condition}. Industrial and experimental chutes often include sidewalls, whose presence is known to substantially modify the \textcolor{black}{flow kinematics}, which in turn would affect segregation as well. The effect of the presence of sidewalls shall be investigated as a separate study in the future.

\section*{Acknowledgments}
AT and SK sincerely acknowledge Indian Institute of Technology Kanpur for funding SK through the Fellowship for Academic and Research Excellence (FA2502005).

\section*{Author Contributions}
Soniya Kumawat and Sayeedul Islam Sheikh contributed equally to the work.

\section*{Notes}
The authors declare no competing financial interest.

\section*{Data Availability Statement}
The data that support the findings of this study are available
from the corresponding author upon reasonable request.

\section*{Supporting information}
\textcolor{black}{Section~\ref{app:Numerical_method} describes the numerical methodology used for the continuum model. Section~\ref{sec:coarse_graining} presents the coarse-graining method used to calculate continuum properties from discrete particle data.}

\printbibliography
\clearpage
\section*{Supplementary Material}
\addcontentsline{toc}{section}{Supplementary Material} 

\setcounter{section}{0}
\renewcommand{\thesection}{SM\arabic{section}}

\renewcommand{\theHsection}{SM\arabic{section}} 

\setcounter{figure}{0}
\renewcommand{\thefigure}{S\arabic{figure}}
\renewcommand{\theHfigure}{S\arabic{figure}}

\setcounter{equation}{0}
\renewcommand{\theequation}{SM\arabic{section}.\arabic{equation}}
\renewcommand{\theHequation}{SM\arabic{section}.\arabic{equation}} 
\makeatletter
\@addtoreset{equation}{section} 
\makeatother

\section{Numerical methodology}
\label{app:Numerical_method}
\hrule
\begin{algorithm}
\caption{Algorithm for predicting the concentration profiles and flow properties}
\textbf{Initialize}:\\
Inlet Layer height $h_{i}$ with spatial grids $N_y = 200$ \\
Length of inclined plane $L_x = 1000 d$, Set $\Delta x = 1d$ \\ 
Species concentration: $f_{i,in}(y)= f_i^T,~0\leq y\leq h_{in}$\\
Solids fraction $\phi(y) = 0.58$ \\
Velocity profile $v_{x,in}(y) = v_0\cdot(y/h_{in})$ and $v_y = v_z = 0$ \\
\While{$x < L_{x}$}
    {
    \textbf{Compute}: \\
    \hspace{2em} Local volume average density using $\rho_{mix}=\displaystyle \sum_{i=1}^{N}\rho_if_{i,in}(y)$  \\
\hspace{2em}Mass flux of $i^{th}$ species $J_{i,in}^{mass}|_{x=0} = \int_0^{H_{in}}\rho_{i}v_{x,in}(y)f_{i,in}(y)dy$\\
    \hspace{2em}Velocity gradient: $|\dot \gamma | = d v_{x}/dy$ numerically\\
    \hspace{2em}Pressure $ P(y) = g\cos{\theta}(1-a\tan{\theta})\int_y^{h(x)}\rho_{b}dy$ numerically \\
    \hspace{2em}Inertial number $I_{mix} = \frac{\dot{\gamma}d_{mix}}{\sqrt{P/\rho_{mix}}}$ \\
    \hspace{2em}Effective friction coefficient $\mu(I_{mix})$ using equation~\ref{eq:mu_I}\\
    Invoke MATLAB PDEPE Solver:\\
     \hspace{2em}Solve simultaneously momentum balance equation (Eq~\ref{eq:vel}) and convection-segregation-diffusion equation (Eq~\ref{eq:conv_seg_diff_den_seg}) along with inlet and boundary conditions (Eq~\ref{eq:conv_seg_diff_bc}) and (Eq~\ref{eq:IC_BC_momentum}), respectively. \\
    Obtain velocity and species concentration profiles, $v_x(y)$ and $f_i(y)$ at $x+\Delta x$\\
    Compute $J_{i}^{mass}|_{x + \Delta x} = \int_0^{h_{in}}\rho_{i}v_{x}(y)f_{i}(y)dy$\\
    Define  error $\epsilon = \sqrt{\sum\limits_{i=1}^N\left| J^{mass}_{i,in}|_{x} - J^{mass}_{i}|_{x+\Delta x}  \right|^2}$ \\
    \While{$\epsilon > 10^{-6}$}
    {
    \eIf{$ J_{i}^{mass}|_{x + \Delta x} > J_{i,in}^{mass}|_{x}  $}{
   $h = h_{in} - \Delta h_0 $\;
   }{
    $h = h_{in} + \Delta h_0 $\;
  }
      Compute $J_{i}^{mass}|_{x + \Delta x} = \int_0^{h} \rho_{i}v_{x}(y)f_{i}(y)dy$ and then compute error $\epsilon$\\
  }
    Update $x\gets x+\Delta x$, $f_{i,in}(y)\gets f_i(y)$ \& $v_{x,in}(y)\gets v_x(y)$ \\
}
\hrule
\label{algo1}
\end{algorithm}

\section{Coarse Graining Method for discrete to continuum properties calculation}
\label{sec:coarse_graining}
The coarse-graining function $W(\vec{r})$ is a semi-definite, positive function, well-defined over the space, and should integrate to unity over the space. To apply the coarse-graining method, particles are assumed to have a convex shape and a single interaction point; contact areas should be replaced by contact points; and collisions should not be instantaneous~\parencite{goldhirsch2010stress}.\\
The spatial scale parameter, $w$, determines the smoothness of macroscopic properties when the coarse-graining method is used. Properties are estimated to be more diffused when a higher $w$ is used, while $w\to0$ would imply concentration of properties at the center of mass and zero everywhere else. For $0.5d\leq w\leq 1d$, the macroscopic field does not show noticeable dependence on the spatial scale. All the properties reported in this work use $w = 0.5d$. Following ref~\cite{weinhart2016influence}, the Gaussian function has been used to define the coarse-graining field in this work to obtain a relatively smoother fields. This coarse-graining function is expressed as:
\begin{equation}
    W(\vec{r}) = \frac{1}{V_w}\exp\left(-\frac{\mid\vec{r}\mid^2}{2w^2}\right) ~\text{if}\mid\vec{r}\mid \leq c~\text{else}~0
    \label{eq:w_r}
\end{equation}
where $V_w$ is the normalization constant and calculated as:
\begin{equation}
    V_w = \int \exp\left(-\frac{\mid\vec{r}\mid^2}{2w^2}\right)d\vec{r}.
    \label{eq:v_w}
\end{equation}
\textcolor{black}{The cutoff length $c$ is chosen to be equal to $1.5d$.} 
For our $3D$ binning case, equation~\ref{eq:v_w} can be solved to obtain the final expression for $W(\vec{r})$ as
\begin{equation}
    W_{3D}(\vec{r}) = \frac{e^{-\frac{|\vec{r}|^2}{2w^2}}}{(2\pi)^{\frac{3}{2}}w^3\left[\text{erf}(\frac{c}{\sqrt{2}w})-\sqrt{\frac{2}{\pi}}\frac{c}{w}e^{-\frac{c^2}{2w^2}}\right]}
    \label{eq:w_r_final}
\end{equation}
The fraction of $i^{th}$ particle on $j^{th}$ cell is computed as:
\begin{equation}
    W(\vec{r}_j-\vec{r}_i) = \frac{W^*(\vec{r}_j-\vec{r}_i)}{\displaystyle \sum_iW^*(\vec{r}_j-\vec{r}_i)}
    \label{eq:cg_w_dem}
\end{equation}
where $W^*(\vec{r}) = \exp{\left(-\frac{|\vec{r}|^2}{2w^2}\right)}$ if $|\vec{r}|<c$ else $0$. The summation used in the denominator allows for direct computation of the normalization constant without computing the integral in equation~\ref{eq:v_w}.

\end{document}